\documentclass[twocolumn,
preprintnumbers,
aps,
floatfix,
prd,
nofootinbib,
superscriptaddress,
showpacs
]{revtex4-1}

\usepackage{amsmath}
\usepackage{amsfonts}
\usepackage{amssymb}
\usepackage{bm}
\usepackage{color}
\usepackage{graphicx} 
\usepackage{subfigure}
\usepackage{multirow, booktabs} 
\usepackage{ragged2e}
\usepackage{float}
\usepackage{hyperref}
\usepackage[all]{hypcap}
\usepackage{ulem}
\usepackage{appendix}
\begin{document}

\title{Extracting the jet transport coefficient from hadron suppressions by confronting current NLO parton fragmentation functions}

\author{Qing-Fei Han}
\affiliation{Key Laboratory of Quark and Lepton Physics (MOE) and Institute of Particle Physics, Central China Normal University, Wuhan 430079, China}

\author{Man Xie}
\affiliation{Key Laboratory of Quark and Lepton Physics (MOE) and Institute of Particle Physics, Central China Normal University, Wuhan 430079, China}

\author{Han-Zhong Zhang}
\affiliation{Key Laboratory of Quark and Lepton Physics (MOE) and Institute of Particle Physics, Central China Normal University, Wuhan 430079, China}
\affiliation{Guangdong Provincial Key Laboratory of Nuclear Science, Institute of Quantum Matter, South China Normal University, Guangzhou 510006, China}
\affiliation{Guangdong-Hong Kong Joint Laboratory of Quantum Matter, Southern Nuclear Science Computing Center, South China Normal University, Guangzhou 510006, China}

\begin{abstract}

Nuclear modification factors of single hadrons and dihadrons at large transverse momentum ($p_{\rm T}$) in high-energy heavy-ion collisions are studied in a next-to-leading-order (NLO) perturbative QCD parton model. Parton fragmentation functions (FFs) in $A+A$ collisions are modified due to jet energy loss which is proportional to the jet transport coefficient $\hat{q}$  characterizing the interaction between the parton jet and the produced medium. By confronting 6 current sets of NLO parton FFs  for large $p_{\rm T}$ hadron productions, we extract $\hat{q}$ quantitatively via a global fit to data for both single hadron and dihadron suppressions, and obtain $\hat{q}/T^3 = 4.74 - 6.72$ at $T = 370$ MeV in central $Au+Au$ collisions at $\sqrt{s_{\rm NN}}=200$ GeV, and $\hat{q}/T^3 = 3.07 - 3.98$ at $T = 480$ MeV in central $Pb+Pb$ collisions at $\sqrt{s_{\rm NN}}=2.76$ TeV. The numerical results show that the uncertainties for $\hat{q}$ extraction are brought by the different contributions of gluon-to-hadron in the 6 sets of FFs due to gluon energy loss being $9/4$ times of quark energy loss.

\vspace{12pt}
\end{abstract}

\maketitle

\section{Introduction}
The hot and dense quark-gluon plasma (QGP) could be produced in high-energy heavy-ion collisions performed at the Relativistic Heavy Ion Collider (RHIC) \cite{Adare:2008qa,Adare:2012wg,Adare:2010sp} and the Large Hadron Collider (LHC) \cite{Abelev:2012hxa,CMS:2012aa,Acharya:2018qsh,Khachatryan:2016odn,Acharya:2018eaq}. When hard partons generated in the initial stage of heavy-ion collisions traverse such a nuclear medium before fragmenting into hadrons, they will lose most of their energy due to multiple scatterings with the strongly-interacting medium. As a consequence, the final state hadrons at large transverse momentum $p_{\rm T}$ are suppressed. This phenomenon is known as Jet Quenching \cite{Gyulassy:1990ye, Wang:1991xy, Qin:2015srf}, which as a hard probe plays an essential role in the studies of QGP properties. The nuclear modification factor, $R_{AA}(p_{\rm T})$ for single hadrons or $I_{AA}(p_{\rm T})$ for dihadrons, is a widely-used and appropriate observable to reflect the suppression effect quantitatively, and is defined as the ratio of the hadron spectrum in nucleus-nucleus ($A+A$) collisions to that in proton-proton ($p+p$) collisions normalized by the number of binary nucleon-nucleon collisions \cite{Zhang:2007ja}.

In recent years, quantitative researches have become the mainstream of a large number of theoretical and experimental studies on jet quenching.
One of the important parts is to extract the jet transport coefficient $\hat{q}$ associated with energy loss by comparing the theoretical results with the experimental data. The parameter $\hat{q}$ characterizes the average transverse momentum broadening squared per unit length for a jet propagating inside the medium \cite{Baier:1996sk},
\begin{equation}
\hat{q}=\rho \int dq_{\rm T}^2\frac{d\sigma}{dq_{\rm T}^2}q_{\rm T}^2,
\label{eq:qhat-define}
\end{equation}
where $\rho$ is the medium gluon number density.
A great amount of abundant experimental data \cite{Adare:2008qa,Adare:2012wg,Adare:2010sp,Abelev:2012hxa,CMS:2012aa,Acharya:2018qsh,Khachatryan:2016odn,Acharya:2018eaq} make it possible to accurately extract jet quenching parameter.
One notable work on the extraction of jet transport coefficient $\hat{q}$ was performed by the JET Collaboration \cite{Burke:2013yra}, which compared several different energy loss models and extracted the $\hat{q}$ values from single hadron suppressions at RHIC and the LHC energies.
Besides, phenomenological investigations have been carried out to extract the initial jet transport coefficient and the mean free path at the initial time simultaneously \cite{Liu:2015vna}, and consider bulk matter evolution \cite{Chen:2010te, Chen:2011vt} for large $p_{\rm T}$ single hadron suppression and other jet quenching observables, such as dihadron and $\gamma$-hadron suppressions \cite{Zhang:2007ja, Zhang:2009rn, Xie:2019oxg, Xie:2020zdb}.
More recently, the jet transport coefficient extractions are continuing via the improving theoretical approaches with opacity-resummed medium induced radiation \cite{Feal:2019xfl}, the quasi-particle collection for QGP based on the linear Boltzmann transport model \cite{Liu:2021dpm}, as well as the advanced analytical technique within JETSCAPE framework \cite{JETSCAPE:2021ehl}.
It is important and necessary for a comprehensive and thorough understanding of the QGP properties to evaluate the systematic uncertainty of jet transport coefficient, which is given by different theoretical methods \cite{Burke:2013yra,Feal:2019xfl,Liu:2021dpm}, observables \cite{Chen:2011vt,Zhang:2007ja, Zhang:2009rn, Xie:2019oxg}, even hydro evolution information \cite{Chen:2010te, Liu:2015vna,Xie:2020zdb}, the initial parton distribution functions and the final fragmentation functions, etc.

A recent investigation for the inclusive charged-particle spectra with a NLO pQCD parton model using seven recent sets of parton-to-hadron fragmentation functions (FFs) indicates that the gluon fragmentation is obviously different in current parton FFs and the hadron production is predominantly sensitive to the gluon-to-hadron FFs in $p+p$ collisions \cite{dEnterria:2013sgr}.
Since gluon energy loss is $9/4$ times quark energy loss \cite{Wang:2009qb}, the use of different fragmentation functions should bring a uncertainty for the jet transport coefficient extracted in high-energy heavy-ion collisions.
Besides, considering that single and dihadorn productions have the same jet energy loss mechanism but slightly different production points for the initial jets \cite{Zhang:2007ja}, the study for the two observables can further help us understand the medium properties. In this paper, we will check the characters of 6 sets of the current NLO parton FFs \cite{Kretzer:2000yf,Kniehl:2000fe,Bourhis:2000gs,Hirai:2007cx,Albino:2008fy,deFlorian:2007aj, deFlorian:2007ekg},
and use a NLO pQCD parton model with jet quenching modified FFs to study high $p_{\rm T}$ hadron productions in heavy-ion collisions \cite{Wang:1996yh,Wang:1996pe,Wang:2004yv,Wang:2009qb,Wang:2001cs,Wang:2002ri}.
Confronting 6 current NLO parton FFs, we will extract the jet transport coefficient $\hat{q}$ quantitatively via a global fit to data for both single hadron and dihadron suppressions and check the extraction uncertainty.

The paper is organized as follows. In Sec. \ref{sec:pqcd-model}, we introduce briefly the NLO pQCD parton model and give the spectrum expressions of single hadron and dihadon productions.
In Sec.\ref{sec:nFFs}, we give comparisons of the 6 current sets of NLO parton FFs variously.
In Sec. \ref{sec:qhat-extracted}, we extract the jet transport coefficient with 6 current sets of FFs by fitting to the experimental data for single hadron and dihadron suppressions at RHIC and the LHC energies. A brief summary and discussions are given in Sec. \ref{sec:summary}.
The above analyses use the same scales in the parton model for different FFs. In the Appendix we add the results and analyses of the $\hat{q}$ extracted from the hadron suppressions with the different scales in the model for different FFs.

\section{The NLO pQCD parton model with modified fragmentation functions} \label{sec:pqcd-model}

\subsection{Cross sections of single hadron and dihadron productions}
According to a NLO pQCD parton model, the differential cross section of single hadron productions in proton-proton ($p+p$) collisions can be obtained as follows \cite{Owens:1986mp,CTEQ:1993hwr},
\begin{eqnarray}
\frac{d\sigma_{pp \rightarrow h+X}}{dyd^2p_{\rm T}}&&=\sum_{abcd}\int dx_a dx_b f_{a/p}(x_a,\mu^2) f_{b/p}(x_b,\mu^2)\nonumber \\
&&\times
\frac{1}{\pi}\frac{d\sigma_{ab\rightarrow cd}}{d\hat{t}}\frac{D_{c}^{h}(z_c,\mu^2)}{z_c}+\mathcal {O}(\alpha_s^3),
\label{eq:pp-sin-spec}
\end{eqnarray}
where $f_{a/p}(x_a,\mu^2)$ is the parton distribution function (PDFs) for a parton $a$ with momentum fraction $x_a$ from a free nucleon, and we will take CT18 parametrizations \cite{Hou:2019efy} in following numerical calculations. $D_{c}^{h}(z_c,\mu^2)$ is the parton FFs in a vacuum, for which we will consider 6 sets of fragmentation function parametrizations in this work. $z_c=p_{\rm T}/{p_{\rm T}}_c$ is the transverse momentum fraction carried by the final hadrons from the parent parton $c$. Here we will focus on hadron productions in the middle-rapidity region. $d\sigma_{ab\rightarrow cd}/d\hat{t}$ is the differential cross section for parton-parton hard scattering process at leading order $\alpha_s^2$. For NLO corrections in $\mathcal {O}(\alpha_s^3)$, we consider both $2\rightarrow3$ real tree diagram contributions and $2\rightarrow2$ one-loop virtual diagram contributions. In our numerical calculations, we use two cutoffs to handle the collinear singularities and soft singularities. The ultraviolet divergences can be solved by renormalization. For more detailed discussions on the NLO calculations, one can find in the references \cite{Kidonakis:2000gi, Harris:2001sx}.

Similarly, the differential cross section of dihadron productions in $p+p$ collisions can be obtained as \cite{Owens:1986mp},\\
\begin{eqnarray}
	\frac{d\sigma_{pp \rightarrow h_1+h_2+X}}{dPS}&&=\sum_{abcd}\int\frac{dz_c}{z_c^2}\frac{dz_d}{z_d^2}f_{a/p}(x_a,\mu^2)f_{a/p}(x_b,\mu^2) \nonumber \\
	&&\times \frac{x_ax_b}{\pi} \frac{d\sigma_{ab\rightarrow cd}}{d\hat{t}}D_{c}^{h_1}(z_c,\mu^2)D_{d}^{h_2}(z_d,\mu^2) \nonumber \\
	&&\times \delta ^2(\frac{\overrightarrow{p}_{\rm T}^{h1}}{z_c}+\frac{\overrightarrow{p}_{\rm T}^{h2}}{z_d})+\mathcal{O}(\alpha_s^3),
\label{eq:pp-di-spec}
\end{eqnarray}
where $dPS=dy^{h_1}d^2p_{\rm T}^{h_1}dy^{h_2}d^2p_{\rm T}^{h_2}$ in the phase space.\\

In high-energy nucleus-nucleus ($A+A$) collisions, the single hadron spectra can be written as \cite{Zhang:2007ja,Zhang:2009rn, Chen:2010te,Liu:2015vna},
\begin{eqnarray}
\frac{dN_{AB \rightarrow h+X}}{dyd^2p_{\rm T}}&&=\sum_{abcd} \int dx_adx_bd^2\vec{r}t_A(\vec{r}) t_B(\vec{r}+\vec{b})  \nonumber \\
&&\times\
f_{a/A}(x_a,\mu^2,\vec{r}) f_{b/B}(x_b,\mu^2,\vec{r}+\vec{b}) \nonumber \\
&&\times \frac{1}{\pi}
\frac{d\sigma_{ab\rightarrow cd}}{d\hat{t}}
\frac{\tilde{D}_{c}^{h}(z_{c},\mu^2,\Delta E_c)}{z_c}+\mathcal {O}(\alpha_s^3).
\label{eq:AA-sin-spec}
\end{eqnarray}

Similarly, the dihadron spectra in $A + A$ collisions can be expressed as \cite{Zhang:2009rn},\\
\begin{eqnarray}
	\frac{dN_{AB \rightarrow h_1+h_2+X}}{dPS}&&=\sum_{abcd}\int \frac{dz_c}{z_c^2}\frac{dz_d}{z_d^2} d^2\vec{r} t_A(\vec{r})t_B(\vec{r}+\vec{b}) \nonumber \\
	&&\times f_{a/A}(x_a,\mu^2,\vec{r})f_{b/B}(x_b,\mu^2,\vec{r}+\vec{b})  \frac{x_ax_b}{\pi} \nonumber \\
	&&\times \frac{d\sigma_{ab\rightarrow cd}}{d\hat{t}} \tilde{D}^{h_1}_{c}(z_c,\mu^2,\Delta E_c)\tilde{D}^{h_2}_{d}(z_d,\mu^2,\Delta E_d) \nonumber \\
	&&\times \delta ^2(\frac{\overrightarrow{p}_{\rm T}^{h1}}{z_c}+\frac{\overrightarrow{p}_{\rm T}^{h2}}{z_d})+\mathcal{O}(\alpha_s^3).
\label{eq:AA-di-spec}
\end{eqnarray}

In Eqs. (\ref{eq:AA-sin-spec}) and (\ref{eq:AA-di-spec}), $t_A(\vec{r})$ is the nuclear thickness function given by the Woods-Saxon distribution \cite{Jacobs:2000wy} for nucleons in a nucleus, and is normalized as $\int d^2\vec{r} t_A(\vec{r}) = A$.
$\vec{b}$ is the impact parameter in $A + A$ collisions.
$f_{a/A}(x_a,\mu^2,\vec{r})$ is the nucleus-modified parton distribution functions, which can be factorized into the parton distribution functions inside a free nucleon $f_{a/N}(x_a,\mu^2)$ and the nuclear shadowing factor $S_{a/A}(x_a,\mu^2,\vec{r})$ \cite{Wang:1996yf,Li:2001xa},
\begin{eqnarray}
f_{a/A}(x_a,\mu^2,\vec{r}) &&= S_{a/A}(x_a,\mu^2,\vec{r})\left[\frac{Z}{A}f_{a/p}(x_a,\mu^2)\right. \nonumber\\
&&+\left.\left(1-\frac{Z}{A}\right)f_{a/n}(x_a,\mu^2)\right],
\end{eqnarray}
where $Z$ is the proton number of the nucleus and $A$ is the nucleus mass number. The nuclear shadowing factor $S_{a/A}(x_a,\mu^2,\vec{r})$ can be obtained using following form \cite{Emelyanov:1999pkc, Hirano:2003pw},
\begin{eqnarray}
S_{a/A}(x_a,\mu^2,\vec{r})&&=1+[S_{a/A}(x_a,\mu^2)-1] \frac{At_A(\vec{r})}{\int{d^2}\vec{r} [t_A(\vec{r})]^2}, \ \ \
\end{eqnarray}
where $S_{a/A}(x_a,\mu^2)$ is given by the EPPS16 parametrizations \cite{Eskola:2016oht}. Since the parton-parton scattering cross sections are computed up to NLO, EPPS16, CT18 and FFs parametrizations are all used at NLO.

$\tilde{D}_{c}^{h}(z_c,\mu^2,\Delta E_c)$  is the medium-modified fragmentation functions and can be calculated as follows \cite{Zhang:2007ja,Zhang:2009rn,Wang:1996yh,Wang:1996pe,Wang:2004yv}:
\begin{eqnarray}
&& \tilde{D}_{c}^{h}(z_c,\mu^2,\Delta{E_c}) = (1-e^{-\langle{N_g}\rangle})\left[\frac{z'_c}{z_c}D_{c}^{h}(z'_c,\mu^2)\right. \nonumber\\
&&\phantom{X}+\left.{\langle{N_g}\rangle}\frac{{z_g}'}{z_c}D_{g}^{h}({z_g}',\mu^2)\right]+e^{-\langle{N_g}\rangle}D_{c}^{h}({z_c},\mu^2),
\label{eq:mffs}
\end{eqnarray}
where ${z_c}'=p_{\rm T}/(p_{\rm{T}c}-\Delta{E_c})$ is the rescaled transverse momentum fraction of the hadron from the quenched parton. The parton has the initial transverse momentum $p_{\rm{T}c}$ to traverse the medium. After losing energy $\Delta{E}_c$, the quenched parton is fragmented into a hadron with momentum $p_{\rm T}$. ${z_g}'= p_{\rm T}/(\Delta{E_c}/\langle{N_g}\rangle)$ is the rescaled transverse momentum fraction of the hadron from the fragmentation of the radiated gluon with the initial energy $\Delta{E_c}/\langle{N_g}\rangle$. $\langle{N_g}\rangle$ is the averaged radiation gluon number and obeys the Poisson distribution. The factor $e^{-\langle{N_g}\rangle}$ is the probability for partons escaping the medium without suffering any inelastic scattering, while the factor $(1-e^{-\langle{N_g}\rangle})$ is the probability for partons encountering at least one inelastic scattering.

$\Delta E_c$ is the total parton energy loss and can be calculated by the high-twist approach \cite{Wang:2009qb,Wang:2001cs,Wang:2002ri}. For a light quark $c$ with the initial energy $E$, the radiative energy loss $\Delta E_c$ can be calculated as,
\begin{eqnarray}
\frac{\Delta{E}_c}{E} &&= \frac{2C_A\alpha_s}{\pi} \int d\tau \int \frac{dl_{\rm T}^2}{l_{\rm T}^4}\int dz \nonumber\\
&&\times \left[1+(1-z)^2\right] \hat{q} \sin^2(\frac{l_{\rm T}^2(\tau-\tau_0)}{4z(1-z)E}),
\label{eq:De}
\end{eqnarray}
where $C_{A}=3$ and $l_{\rm T}^2$ is the squared transverse momentum of the radiated gluon. Since the colour factor of gluon-gluon vertex is  $9/4$ times that of quark-gluon vertex, the energy loss of a gluon jet is $9/4$ times that of a quark jet \cite{Wang:2009qb}.
The average number of the gluons emitted off a hard parton is calculated as \cite{Chang:2014fba},
\begin{eqnarray}
\langle N_g \rangle &&= \frac{2C_A \alpha_{s}}{\pi} \int d\tau \int \frac{dl_{\rm T}^2}{l_{\rm T}^4}\int \frac{dz}{z} \nonumber\\
&&\times \left[1+(1-z)^2\right] \hat{q} \sin^2(\frac{l_{\rm T}^2(\tau-\tau_0)}{4z(1-z)E}).
\label{eq:Ng}
\end{eqnarray}

\subsection{The jet transport coefficient}

The total parton energy loss and the number of radiated gluons are both controlled by jet transport coefficient $\hat{q}$ \citep{Baier:1996sk}. According to Eq. (\ref{eq:qhat-define}) for the definition of jet transport coefficient, one can assume the $\hat{q}$ is proportional to the local gluon density in a QGP phase or the hadron density in a hadronic gas. Therefore the transport coefficient in an evolving dynamical medium can be expressed as \cite{Chen:2010te, Chen:2011vt},
\begin{eqnarray}
 \hat{q}(\tau,\vec{r}) &&= \left[ \hat{q}_0\frac{\rho_{\scriptscriptstyle QGP}(\tau,\vec{r}+(\tau-\tau_0)\vec{n})}{\rho_{\scriptscriptstyle QGP}(\tau_0,0)}(1-f) \right. \nonumber\\
&&+ \left. {\hat{q}_{\rm had}(\tau,\vec{r}+(\tau-\tau_0)\vec{n})f} \frac{ }{ } \right] \cdot\frac{p^{\mu}u_{\mu}}{p_0},
\label{eq:qhat}
\end{eqnarray}
where $\vec{n}$ is the unit length vector in the parton jet moving direction.
The first part in the above equation represents the contribution in the QGP phase, while the second part denotes in the hadronic phase.
We also consider the effective flow dependence of jet transport coefficient. $p^{\mu}$ is the four-momentum of the jet, and $u_{\mu}$ is the four flow velocity of the medium. $f(\tau,\vec{r})$ is the hadronic phase fraction at a given time-space point, which can distinguish the contribution of jet energy loss in the QGP and the hadronic phase,
\begin{equation}\label{f-fraction}
f(\tau,\vec{r})=\begin{cases}
0 & T>T_c \\
0\sim 1 & T=T_c \\
1 & T<T_c.
\end{cases}
\end{equation}
Here $T$ is the local temperature of the medium. In our studies, the hydrodynamical time-space evolution information of medium temperature $T$ and flow velocity $u$ are obtained by the (3+1)-dimensional ideal hydrodynamic model \cite{Hirano:2001eu,Hirano:2002ds}, in which there is a mixed phase platform of the first order phase transition between the QGP and hadron phases at the critical temperature  $T_c=170$ MeV.

For the QGP phase in Eq. \ref{eq:qhat}, $\hat{q}_0$ denotes the jet transport coefficient at the center of the bulk medium in the initial time $\tau_0$. $\rho_{\scriptscriptstyle QGP}$ is the parton density at a given time and space, which is propotional to the temperature cubed \cite{Hirano:2001eu,Hirano:2002ds,Chen:2010te, Chen:2011vt}. Thereby in numerical caculations, we assume that $\hat{q}$ for the QGP phase has the following form as \cite{Liu:2015vna, Xie:2019oxg},
\begin{eqnarray}
\hat q =\hat q_0 \frac{T^3}{T_0^3} \frac{p^{\mu}u_{\mu}}{p_0}(1-f),
\label{eq:qhat-cons}
\end{eqnarray}
where $T_0$ is a reference temperature taken as the highest temperature in the center of the medium at the initial time $\tau_0$.

\begin{table*}[htbp]
\centering
\caption{\label{table1} Characteristics of the six sets of current fragmentation function parameterizations with hadron species, fitted data, minimum value of $z$, and the scale range of $\mu^2$.}
\begin{ruledtabular}
\begin{tabular}{lllll} 
{\small FFs set} & {\small Species} & {\small Fitted data} & $z_{min}$ & $\mu^2$(GeV$^2$)\\
\hline
KRE \cite{Kretzer:2000yf} & $\pi^{+}, \pi^{-}, K^{+}, K^{-}, h^{+}, h^{-}$ & $e^+e^-$ & 0.01 & $0.8-10^6$\\
KKP \cite{Kniehl:2000fe} & $\pi^{+}+\pi^{-}$, $K^{+}+K^{-}$, $p+\overline{p}$, $h^{+}+h^{-}$  & $e^+e^-$ & 0.1 & $2.0-4.0\cdot 10^4$ \\
BFGW \cite{Bourhis:2000gs} & $h^{+}+h^{-}$ & $e^+e^-$ & 0.001 & $2.0-1.2\cdot 10^4$ \\
HKNS \cite{Hirai:2007cx} & $\pi^{+}, \pi^{-}, K^{+}, K^{-}, p, \overline{p}, h^{+}, h^{-}$ & $e^+e^-$ & 0.01 & $1.0-10^8$ \\
AKK08 \cite{Albino:2008fy} & $\pi^{+}, \pi^{-}, K^{+}, K^{-}, p, \overline{p}, h^{+}, h^{-}$ & $e^+e^-$, $pp$ & 0.05 & $1.0-10^6$ \\
DSS \cite{deFlorian:2007aj, deFlorian:2007ekg} & $\pi^{+}, \pi^{-}, K^{+}, K^{-}, p, \overline{p}, h^{+}, h^{-}$ & $e^+e^-$, $pp$, $ep$ & 0.05 & $1.0-10^5$ \\
\end{tabular}
\end{ruledtabular}
\end{table*}

For the hadronic phase in Eq. \ref{eq:qhat}, $\hat{q}_{\rm had}$ is the jet transport coefficient in the hadronic phase and can be written as \cite{Chen:2010te},
\begin{eqnarray}
\hat{q}_{\rm had}=\frac{\hat{q}_N}{\rho_N}[\frac{2}{3}\sum\limits_{M}\rho_M(T)+\sum\limits_{B}\rho_B(T)],
\label{eq:qhat-had}
\end{eqnarray}
where $\hat{q}_N$ is the extracted jet transport coefficient at the center of a large nucleus and given by $\hat{q}_N\approx 0.02$ GeV$^2/$fm, and $\rho_N \approx 0.17$ fm$^{-3}$ is the nucleon density at the center of the large nucleus \cite{Chen:2010te}. $\rho_M$ and $\rho_B$ are the meson and baryon density in the hadronic resonance gas at a given temperature, respectively.
The factor 2/3 denotes the ratio of constituent quark numbers in mesons and baryons.
The hadron density at a temperature $T$ and zero chemical potential is expressed as \cite{Chen:2010te},
\begin{eqnarray}
\sum\limits_{h}\rho_h(T)=\frac{T^3}{2\pi^2}\sum\limits_{h}(\frac{m_h}{T})^2\sum\limits_{n=1}^{\infty}\frac{\eta_h^{n+1}}{n}K_2(n\frac{m_h}{T}),
\end{eqnarray}
where $\eta_h=\pm 1$ for meson (M)/baryon (B). In this paper, we will consider hadron resonances, and the mass is below 1 GeV, including all 17 kinds of mesons: $\pi^+$, $\pi^-$, $\pi^0$, $K^+$, $K^-$, $K^0$, $\overline{K^{0}}$, $\eta$, $\eta^{'}$, $\rho^+$, $\rho^-$, $\rho^0$, $K^{*+}$, $K^{*-}$, $K^{*0}$, $\overline{K^{*0}}$, $\omega$; and 2 kinds of baryons: $p$, $n$.

\subsection{Nuclear modification factors}

Confronting the 6 current sets of parton fragmentation functions, we will extract the parameter $\hat{q}_0$ using the $\chi^2$ fitting method of comparing the NLO pQCD numerical results of single hadron and dihadron suppressions with the experimental data.
To demonstrate the suppression of the single hadron spectrum in $A+A$ collisions relative to that in $p+p$ collisions, one can define the nuclear modification factor $R_{AA}(p_{\rm T})$ as \cite{Wang:2004yv},
\begin{eqnarray}
R_{AA}(p_{\rm T})=\frac{1}{T_{AA} (\vec{b})}\frac{dN_{AA\rightarrow h+X}/dyd^2p_{\rm T}}{d{\sigma}_{pp\rightarrow h+X}/dyd^2p_{\rm T}},
\label{eq:Raa}
\end{eqnarray}
where $T_{AA}(\vec{b}) =\int d^2\vec{r} t_A(\vec{r})t_A(\vec{r}+\vec{b})$ is the overlap function of two colliding nuclei for a given impact parameter.

In the following numerical studies for dihadrons, for a given triggered hadron, we will focus on the away-side associated hadrons.
The nuclear modification factor $I_{AA}$ for dihadron productions can be defined as a function of $z_{\rm T}=p_{\rm T}^{\rm assoc}/p_{\rm T}^{\rm trig}$ or a function of $p_{\rm T}^{\rm assoc}$,
\begin{equation}
   I_{AA}(z_{\rm T})=\frac{D_{AA}(z_{\rm T})}{D_{pp}(z_{\rm T})},
\label{eq:Iaa-zt}
\end{equation}
or
\begin{equation}
   I_{AA}(p_{\rm T}^{\rm assoc})=\frac{D_{AA}(p_{\rm T}^{\rm assoc})}{D_{pp}(p_{\rm T}^{\rm assoc})},
\label{eq:Iaa-pt}
\end{equation}
where $D_{AA}(z_{\rm T})=p_{\rm T}^{\rm trig}D_{AA}(p_{\rm T}^{\rm assoc})$ is called hadron-triggered fragmentation function, and can be calculated by \cite{Wang:2003aw},
\begin{equation}
   D_{AA}(z_{\rm T}) = p_{\rm T}^{h_1}\frac{dN_{AA\rightarrow h_1+h_2+X}/dy^{h_1}dp_{\rm T}^{h_1}dy^{h_2}dp_{\rm T}^{h_2}}{dN_{AA\rightarrow h_1+X}/dy^{h_1}dp_{\rm T}^{h_1}}.
 \label{eq:DAB}
\end{equation}

We calculate the $R_{AA}$ for single hadrons and $I_{AA}$ for dihadrons with the six sets of FFs at RHIC and the LHC energies, and compare the numerical results to the experimental data by utilizing $\chi^2/d.o.f$ fitting method. The $\chi^2/d.o.f$ is defined as follows,
\begin{equation}
\chi^2/d.o.f=\sum_{i=1}^{N}\left[\frac{(V_{th}-V_{exp})^2}{\sigma_{sys}^2+ \sigma_{stat}^2}\right]_i /N
\label{eq:chi2}
\end{equation}
where $V_{th}$ represents the theoretical value, $V_{exp}$ denotes the experimental data,  $\sigma_{sys}$ and $\sigma_{stat}$ are the systematic and statistical errors for the experimental data, and $N$ is the number of data points which are used.

\section{Comparisons of parton fragmentation functions} \label{sec:nFFs}

In this work, we employ the six sets of commonly-used parameterizations of NLO parton-to-hadron fragmentation functions: Kretzer (KRE) \cite{Kretzer:2000yf}, KKP \cite{Kniehl:2000fe}, BFGW \cite{Bourhis:2000gs}, HKNS \cite{Hirai:2007cx}, AKK08 \cite{Albino:2008fy}, and DSS \cite{deFlorian:2007aj, deFlorian:2007ekg}, as listed in Table \ref{table1}. From the species column, we can see that the charged hadrons ($h^{\pm}$) are constructed as a sum of the individual FFs for pions ($\pi^{\pm}$), kaons ($K^{\pm}$), and (anti)protons($p, \overline{p}$) in most of the parton FFs sets except the BFGW FFs.
$\pi^0$ hadron is given by the average of pions ($\pi^{+}$ and $\pi^{-}$).
In this section we will briefly show the differences of parton-to-$\pi^0$ or parton-to-$h^{\pm}$ between these six sets of fragmentation function parameterizations.

\begin{figure}[htbp]
\includegraphics[width=0.38\textwidth]{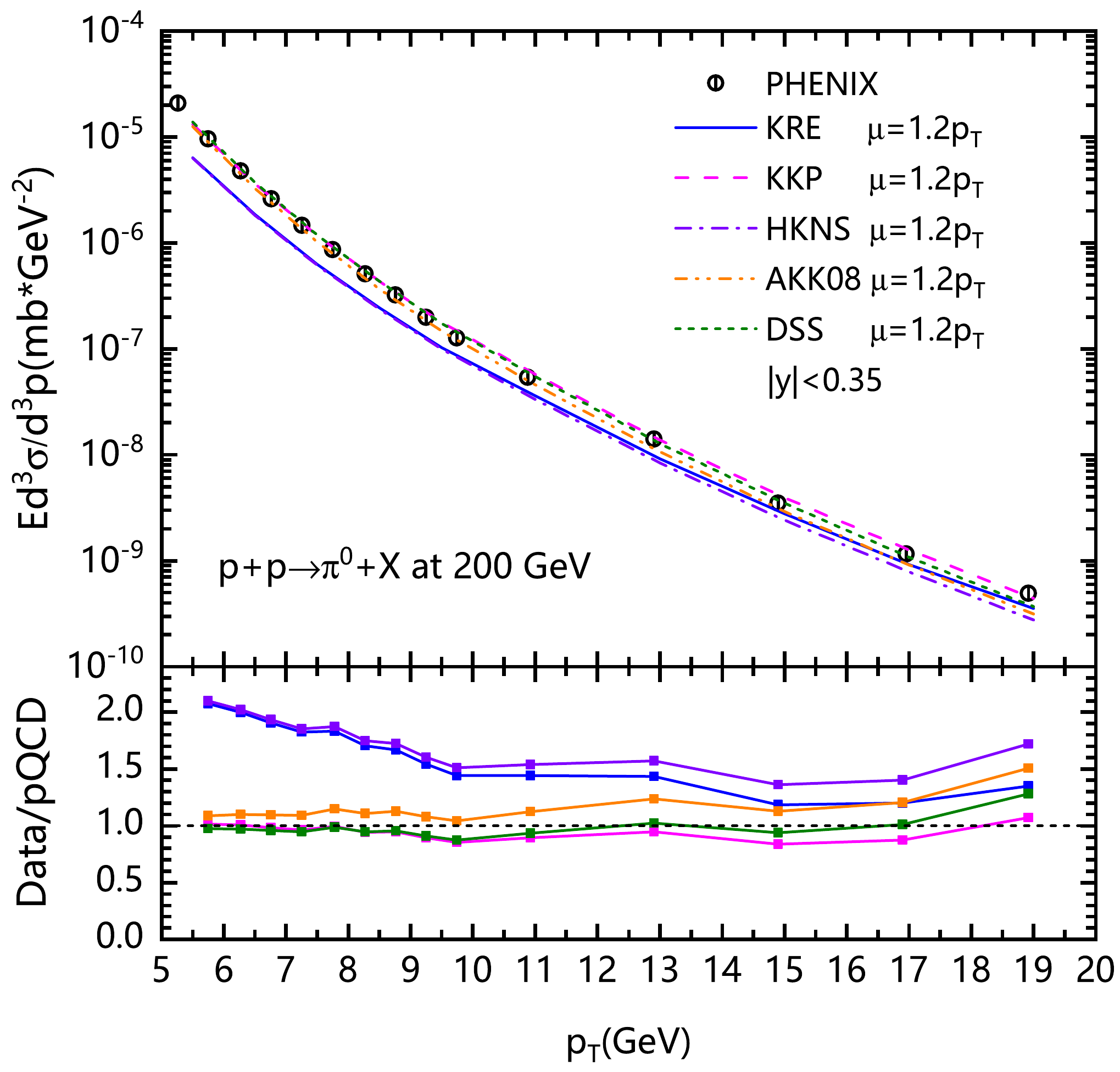}
\caption{The invariant cross sections of single $\pi^0$ for the same scale $\mu=1.2p_{\rm T}$ with different FFs in $p+p$ collisions at $\sqrt{s_{\rm NN}}=200$ GeV (upper panel), and ratios of data over the theoretical results (lower panel). The data are from Ref. \cite{Adare:2008qa}. }
\label{fig:pp_rhic_s}
\end{figure}

\begin{figure}[htbp]
\includegraphics[width=0.42\textwidth]{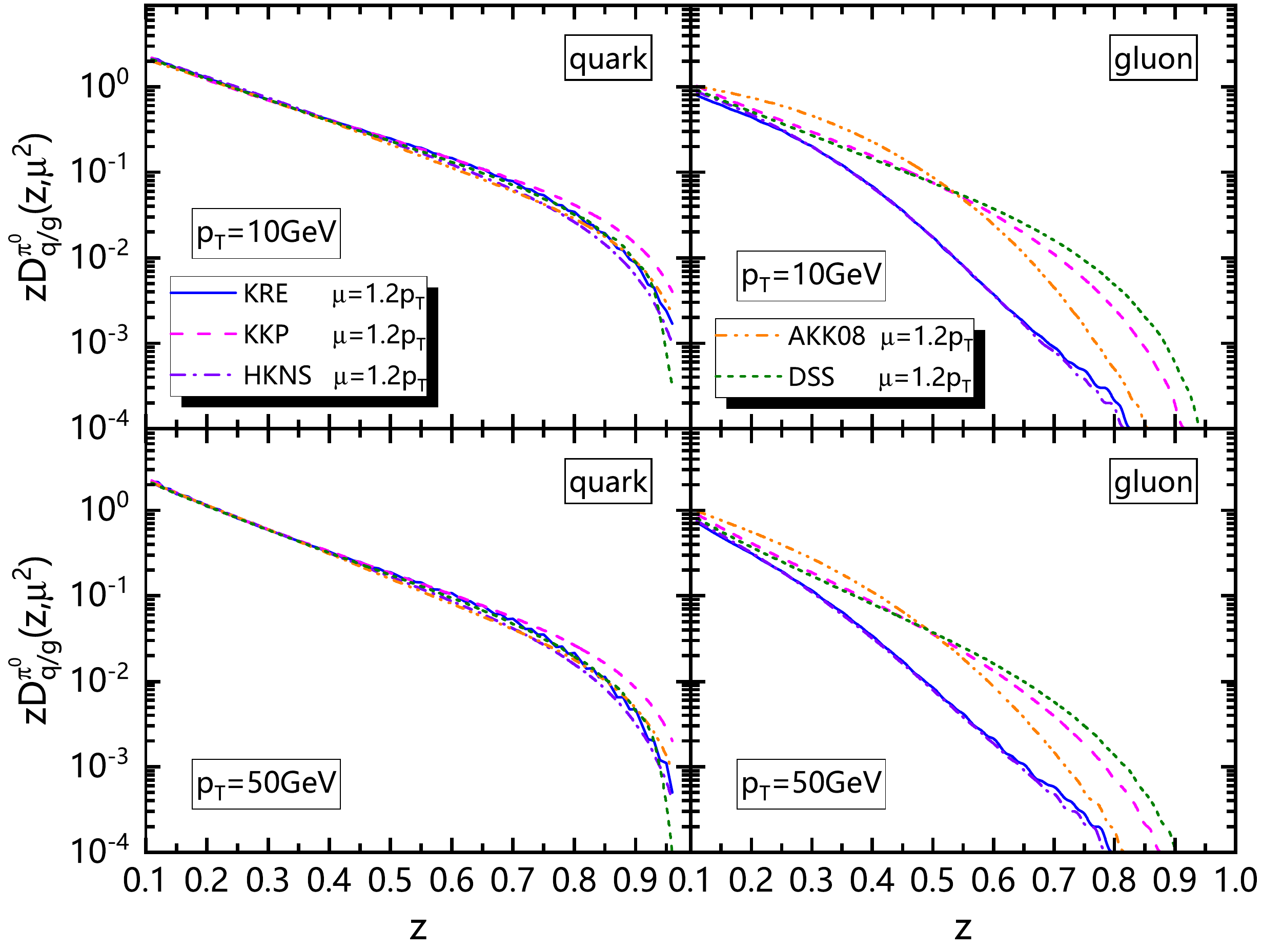}
\caption{Comparisons of parton-to-$\pi^0$ fragmentation functions between 5 sets of FFs, KRE\cite{Kretzer:2000yf}, KKP \cite{Kniehl:2000fe}, HKNS \cite{Hirai:2007cx}, AKK08 \cite{Albino:2008fy}, and DSS \cite{deFlorian:2007aj, deFlorian:2007ekg}. Left panels are for quark FFs, and right panels for gluon FFs. Upper panels are for hadrons with $p_{\rm T}=10$ GeV, and lower panels with $p_{\rm T}=50$ GeV. The scale is set as $\mu = 1.2 p_{\rm T}$.}
\label{fig:zD-z-pi_rhic_s}
\end{figure}

\begin{figure}[htbp]
\includegraphics[width=0.42\textwidth]{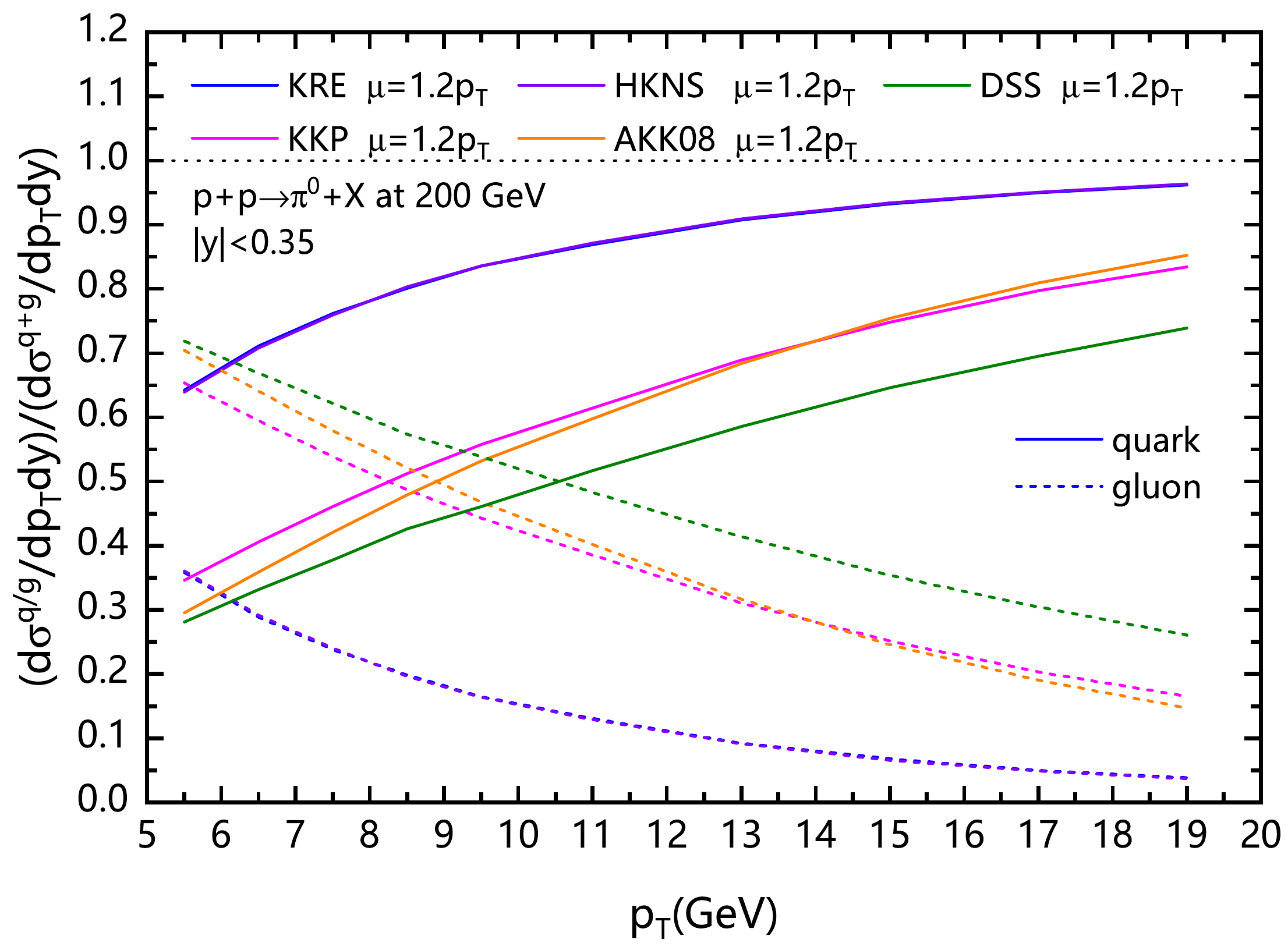}
\caption{The contribution fractions of quark (solid lines) and gluon (dashed lines) fragmentations to the inclusive $\pi^0$ cross sections for different FFs in $p+p$ collisions at $\sqrt{s_{\rm NN}}=200$ GeV.}
\label{fig:frac_rhic_s}
\end{figure}

\subsection{Characteristics of parton FFs for $\pi^0$ hadrons}

In the NLO pQCD parton model for hadron productions, there are three independent scales: the factorization scale $\mu_{\rm fac}$, the renormalization scale $\mu_{\rm ren}$ and the fragmentation scale $\mu_{\rm fra}$. In following numerical calculations we choose the equal scales $\mu_{\rm fac}=\mu_{\rm ren}=\mu_{\rm fra}$, and let them proportional to the physics observables, such as hadron transverse momentum $p_{\rm T}$ or dihadron invariant mass $M$.

As shown in the upper panel of Fig.~\ref{fig:pp_rhic_s}, we firstly give the NLO pQCD results of differential cross sections for $\pi^0$ productions at large transverse momentum $p_{\rm T}$ with the 5 sets of FFs in $p+p$ collisions at $\sqrt{s_{\rm NN}}=200$ GeV, and compare them with the experimental data \cite{Adare:2008qa}. Here we choose the scale as $\mu = 1.2 p_{\rm T}$. The numerical results can describe the experimental data. There are certain differences in the specific results from the different fragmentation function parameterizations, as shown in the lower panel of Fig.~\ref{fig:pp_rhic_s} for the ratios of the experimental data to theoretical calculations.

Note that such differences mainly result from the discrepancies of gluon-to-$\pi^0$ FFs \cite{dEnterria:2013sgr}.
We give in Fig.~\ref{fig:zD-z-pi_rhic_s} the comparisons of the quark (including u, d, and s) and gluon FFs for $\pi^0$ hadrons at $p_{\rm T}=10$ GeV and 50 GeV, respectively. The FFs of quark-to-$\pi^0$ give similar contributions, while the gluon-to-$\pi^0$ FFs provide significantly different contributions during the whole range of $z$. For examples, since KRE and HKNS parameterizations have less gluon contributions relative to other sets, the model with KRE or HKNS FFs underestimates $\pi^0$ productions in $p+p$ collisions at 200 GeV as shown in Fig.~\ref{fig:pp_rhic_s}. Of course, in the NLO pQCD parton model using KRE and HKNS fragmentation function parameterizations, one can add a $K$ factor ($K > 1.0$) to account for higher-order contributions or choose a smaller scale $\mu$ to fit data better for the $p_{\rm T}$ spectra in $p+p$ collisions.

To make clear the relative contributions of quarks and gluons to final state hadrons in these sets of FFs, we show in Fig.~\ref{fig:frac_rhic_s} the contribution fractions of quarks (solid lines) or gluons (dashed lines) to the inclusive charged-hadron cross sections in $p+p$ collisions. One can see that with the same scale $\mu=1.2p_{\rm T}$, the relative contributions of quarks and gluons to $\pi^0$ productions are both distinctly different for all the sets of FFs due to the apparent differences in gluon fragmentation functions.

\begin{figure}[htbp]
\includegraphics[width=0.38\textwidth]{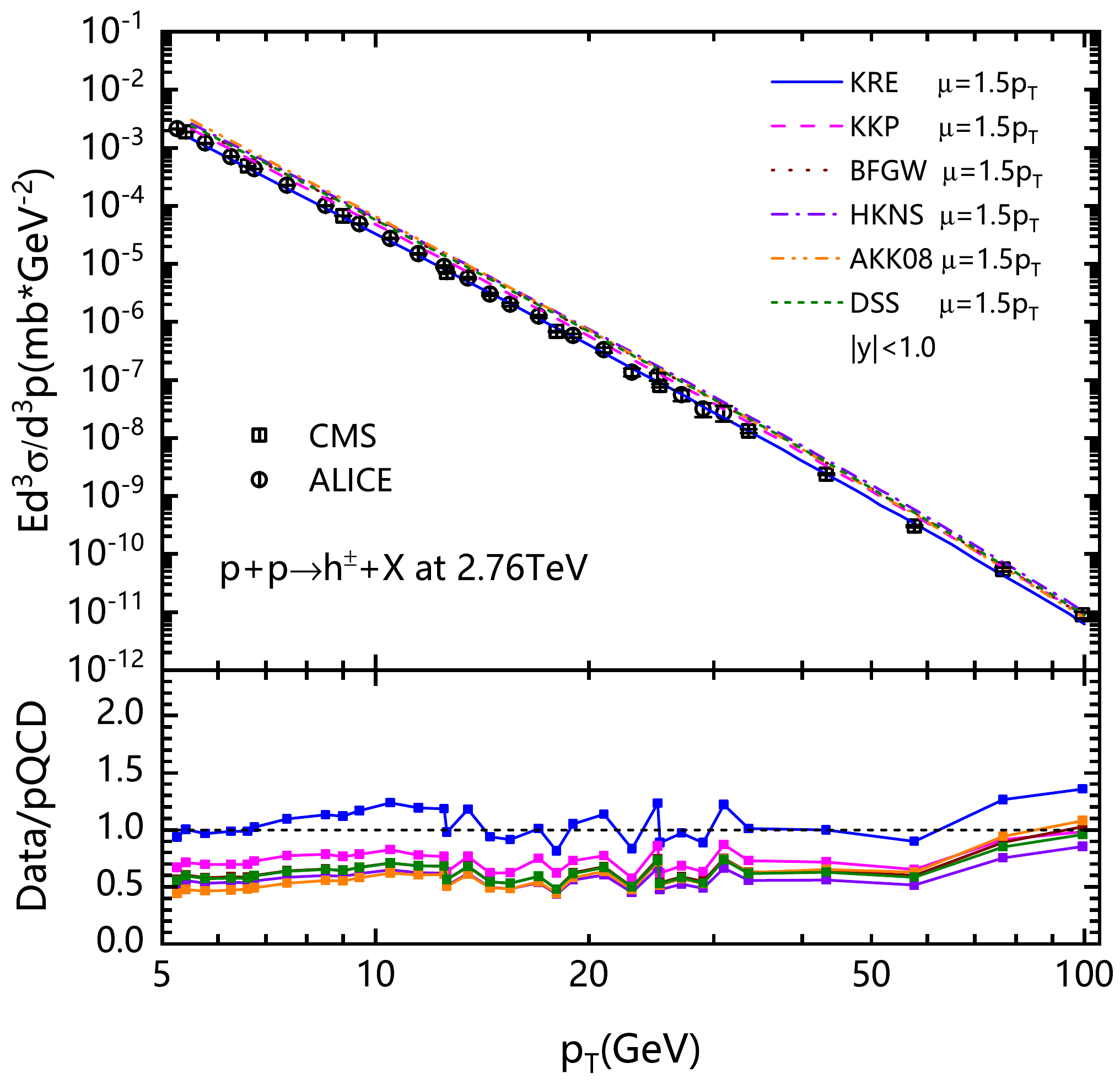}
\caption{The invariant cross sections of single charged hadrons for the same scale $\mu=1.5p_{\rm T}$ with different FFs in $p+p$ collisions at $\sqrt{s_{\rm NN}}=2.76$ TeV (upper panel), and ratios of data over the theoretical results (lower panel). The data are from Refs. \cite{CMS:2012aa, Abelev:2012hxa}.}
\label{fig:pp_lhc_s}
\end{figure}

\begin{figure}[htbp]
\includegraphics[width=0.42\textwidth]{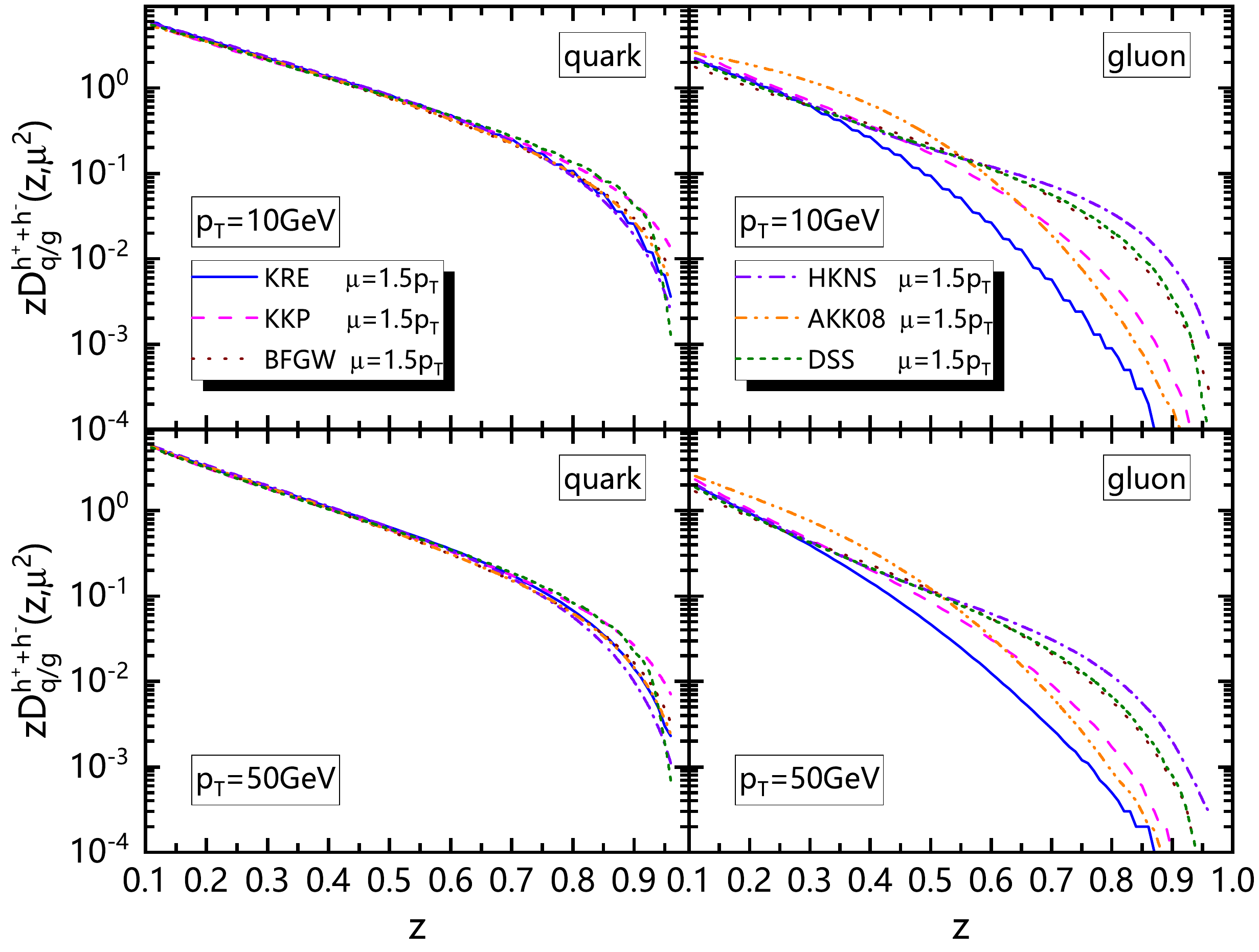}
\caption{Similar to Fig. \ref{fig:zD-z-pi_rhic_s} but 6 sets of FFs for charged hadrons. The scale is set as $\mu = 1.5 p_{\rm T}$.}
\label{fig:zD-z-h_lhc_s}
\end{figure}

\begin{figure}[htbp]
\includegraphics[width=0.42\textwidth]{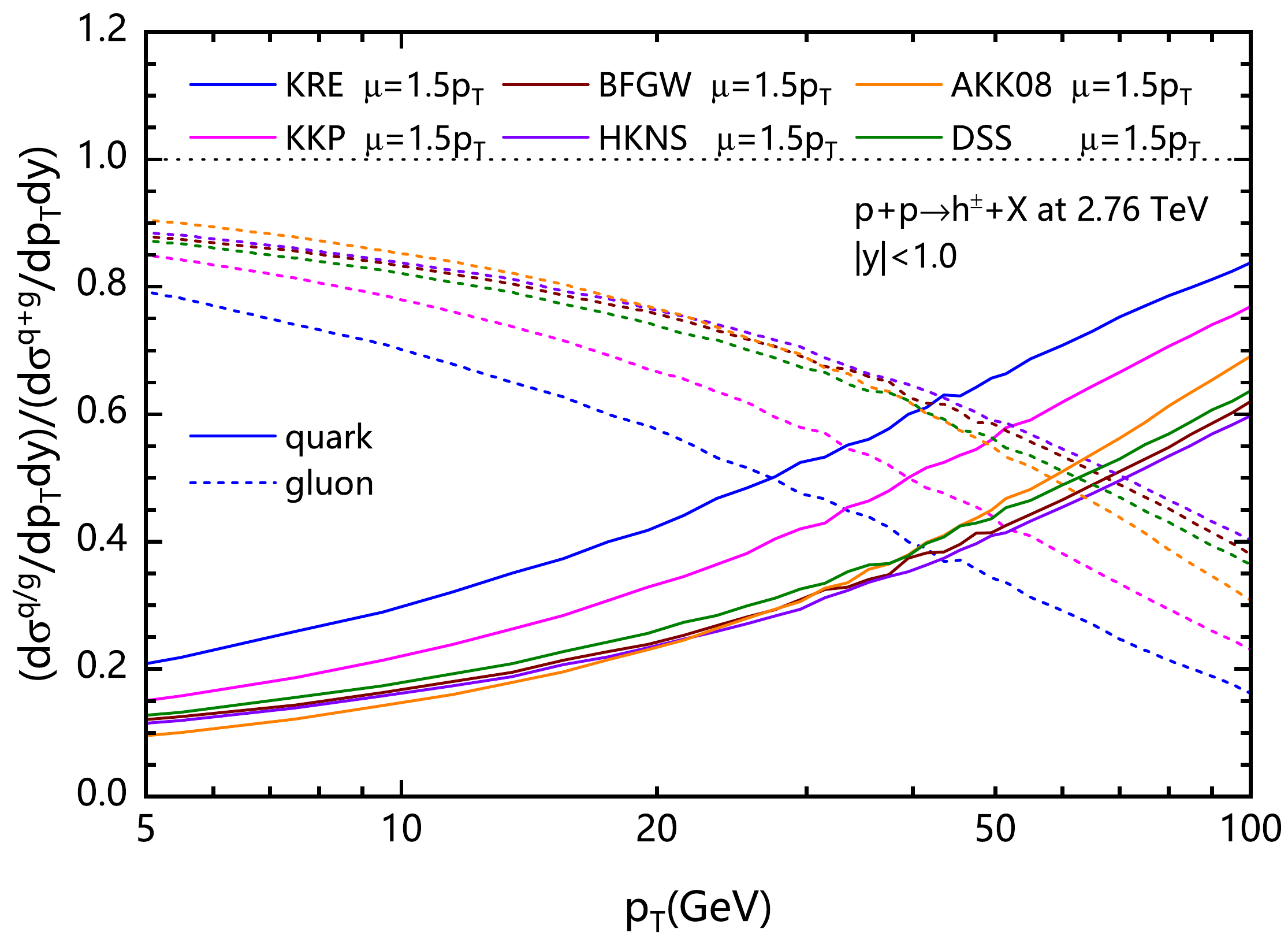}
\caption{The contribution fractions of quark (solid lines) and gluon (dashed lines) fragmentations to the inclusive charged-hadron cross sections for different FFs in $p+p$ collisions at $\sqrt{s_{\rm NN}}=2.76$ TeV.}
\label{fig:frac_lhc_s}
\end{figure}

\subsection{Characteristics of parton FFs for charged hadrons}

Similar to the study at RHIC energy, we also calculate the cross sections of charged hadrons in $p+p$ collisions at $\sqrt{s_{\rm NN}}=2.76$ TeV, and give ratios of the experimental data to theoretical results, as shown in Fig.~\ref{fig:pp_lhc_s}. We choose $\mu = 1.5p_{\rm T}$ as the default scale for charged hadron productions. The numerical results for charged hadrons with different FFs can also describe the experimental data, but the model with the most of the FFs overestimates slightly the charged hadron productions with $\mu = 1.5p_{\rm T}$.

Fig.~\ref{fig:zD-z-h_lhc_s} shows the comparisons for the quark and gluon FFs of charged hadrons at $p_{\rm T}=10$ GeV (upper panels) and 50 GeV (lower panels) from the 6 sets of FFs, respectively. Likewise to pion productions, the FFs of quark-to-$h^{\pm}$ give similar contributions, but the gluon-to-$h^{\pm}$ FFs show significantly different contributions during the whole range of $z$.
Compared to KRE gluon fragmentation, the too-hard gluon contributions of the other sets of FFs lead to the overestimation in Fig.~\ref{fig:pp_lhc_s} for the charged hadron productions in $p+p$ collisions at $\sqrt{s_{\rm NN}}=2.76$ TeV.

For a more intuitive view, the relative contributions from the quark and gluon fragmentations are given in Fig.~\ref{fig:frac_lhc_s} for the single charged hadrons in $p+p$ collisions at $\sqrt{s_{\rm NN}}=2.76$ TeV.
Similarly to Fig.~\ref{fig:frac_rhic_s}, there exists different fractions of gluon (quark) contributions to hadrons among the 6 sets of FFs.

\section{Jet transport coefficient extractions} \label{sec:qhat-extracted}

Because gluon energy loss is 9/4 times of quark energy loss, the different fractions of gluon (quark) contributions to hadrons in the model with different sets of FFs will give different strength of jet quenching for high $p_{\rm T}$ hadron productions in high-energy nucleus-nucleus collisions.
In this section, confronting the difference in the six sets of FFs, let's check the jet quenching parameter $\hat{q}_0$ extracted from both the single hadron and dihadron suppressions by fitting data at RHIC and the LHC.

\subsection{Extracting $\hat{q}_0$ in central $Au+Au$ collisions at $\sqrt{s_{\rm NN}}=200$ GeV}

We choose $b=2.0$ (3.2) fm in Eq. (\ref{eq:Raa}) for $R_{AA}(p_{\rm T})$ and Eq. (\ref{eq:Iaa-zt}) for $I_{AA}(z_{\rm T})$ in 0-5\% (0-10\%) Au+Au collisions. Shown in Fig.~\ref{fig:raa_rhic_s} are the nuclear modification factors $R_{AA}(p_{\rm T})$ for single hadrons (left panels) in $0-5\%$ $Au+Au$ collisions,  and $I_{AA}(z_{\rm T})$ for dihadrons (right panels) in $0-10\%$ $Au+Au$ collisions at $\sqrt{s_{\rm NN}}=200$ GeV. KRE\cite{Kretzer:2000yf}, KKP \cite{Kniehl:2000fe}, HKNS \cite{Hirai:2007cx}, AKK08 \cite{Albino:2008fy}, and DSS \cite{deFlorian:2007aj, deFlorian:2007ekg} FFs are used with the same scale $\mu = 1.2 p_{\rm T}$ for single hadrons and $\mu = 1.2 M$ for dihadrons, respectively. Theoretical results fit data well with several appropriate values of jet transport coefficient $\hat{q}_0$. The solid curves are given by the best fitting from the next $\chi^2$-fitting calculations.

Shown in Fig.~\ref{fig:kai2_rhic_s} are the $\chi^2/d.o.f$ fits to nuclear modification factors in central $Au+Au$ collisions at $\sqrt{s_{\rm NN}}=200$ GeV, (a) the fits to only single hadron $R_{AA}(p_{\rm T})$, (b) the fits to only dihadron $I_{AA}(z_{\rm T})$, and (c) the global fits to $R_{AA}(p_{\rm T})$ + $I_{AA}(z_{\rm T})$. The 5 sets of fragmentation function parameterizations are used respectively. $p_{\rm T} > 5$ GeV for single hadrons $R_{AA}(p_{\rm T})$ and $z_{\rm T} > 0.25$ for $\pi^0$-triggered away-side charged hadrons $I_{AA}(z_{\rm T})$ are chosen to fit data.

The same mechanism of jet quenching leads to both suppressions of large $p_{\rm T}$ single hadrons and $\pi^0$-triggered away-side charged hadrons in heavy-ion collisions, so the best-fitting values of $\hat{q}_0$ for single hadrons in panel (a) of Fig.~\ref{fig:kai2_rhic_s} are similar to those for dihadrons in panel (b) for most sets of FFs used in the theoretical model.
From a global $\chi^2$ fits to both the single hadron and dihadron suppressions in panel (c) of Fig.~\ref{fig:kai2_rhic_s}, the best-fitting values of $\hat{q}_0$ for each set of FFs are obtained as: $\hat{q}_0=1.7$ GeV$^2$/fm with KRE FFs, $\hat{q}_0=1.4$ GeV$^2$/fm with KKP FFs, $\hat{q}_0=1.7$ GeV$^2$/fm with HKNS FFs, $\hat{q}_0=1.3$ GeV$^2$/fm with AKK08 FFs, and $\hat{q}_0=1.2$ GeV$^2$/fm with DSS FFs.

\begin{widetext}

\begin{figure}[htbp]
\includegraphics[width=0.38\textwidth]{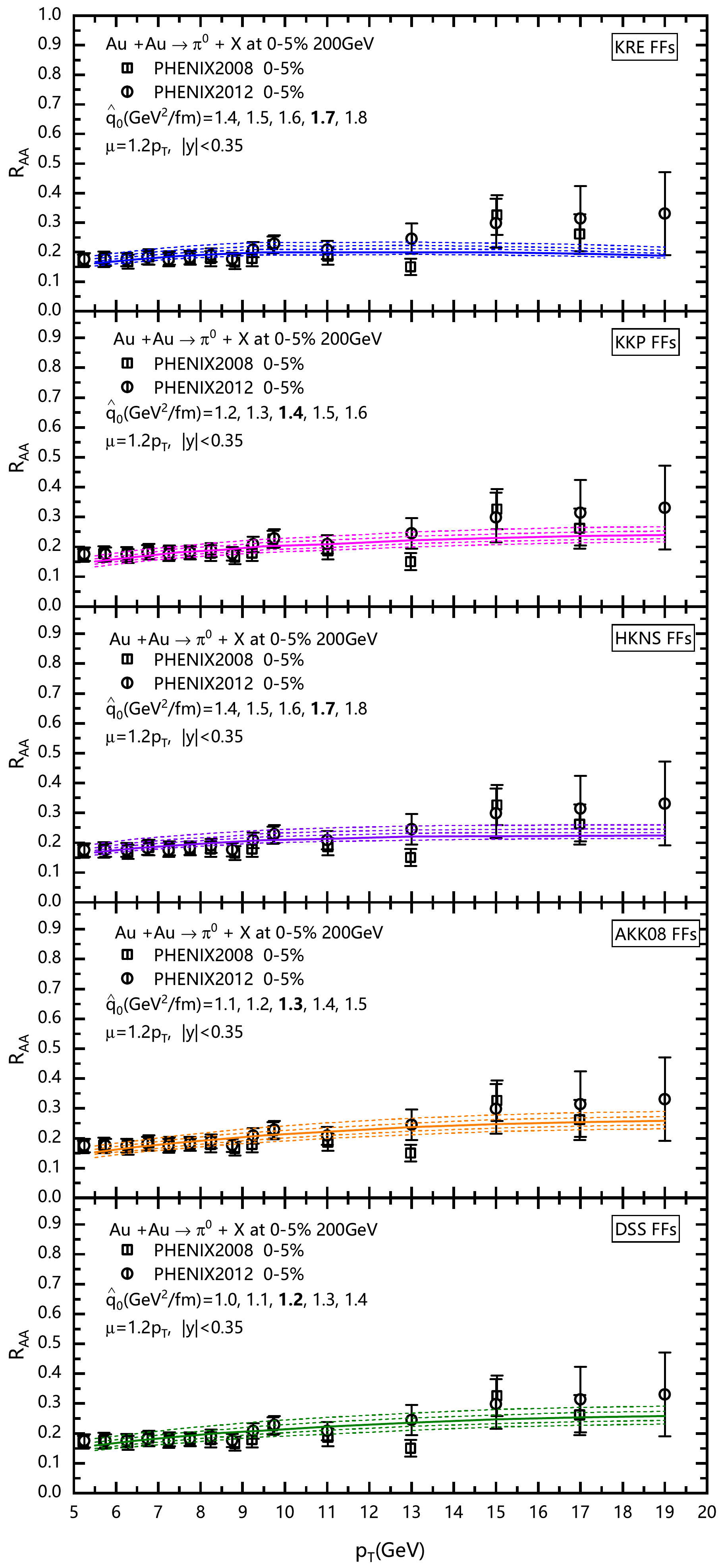}
\includegraphics[width=0.38\textwidth]{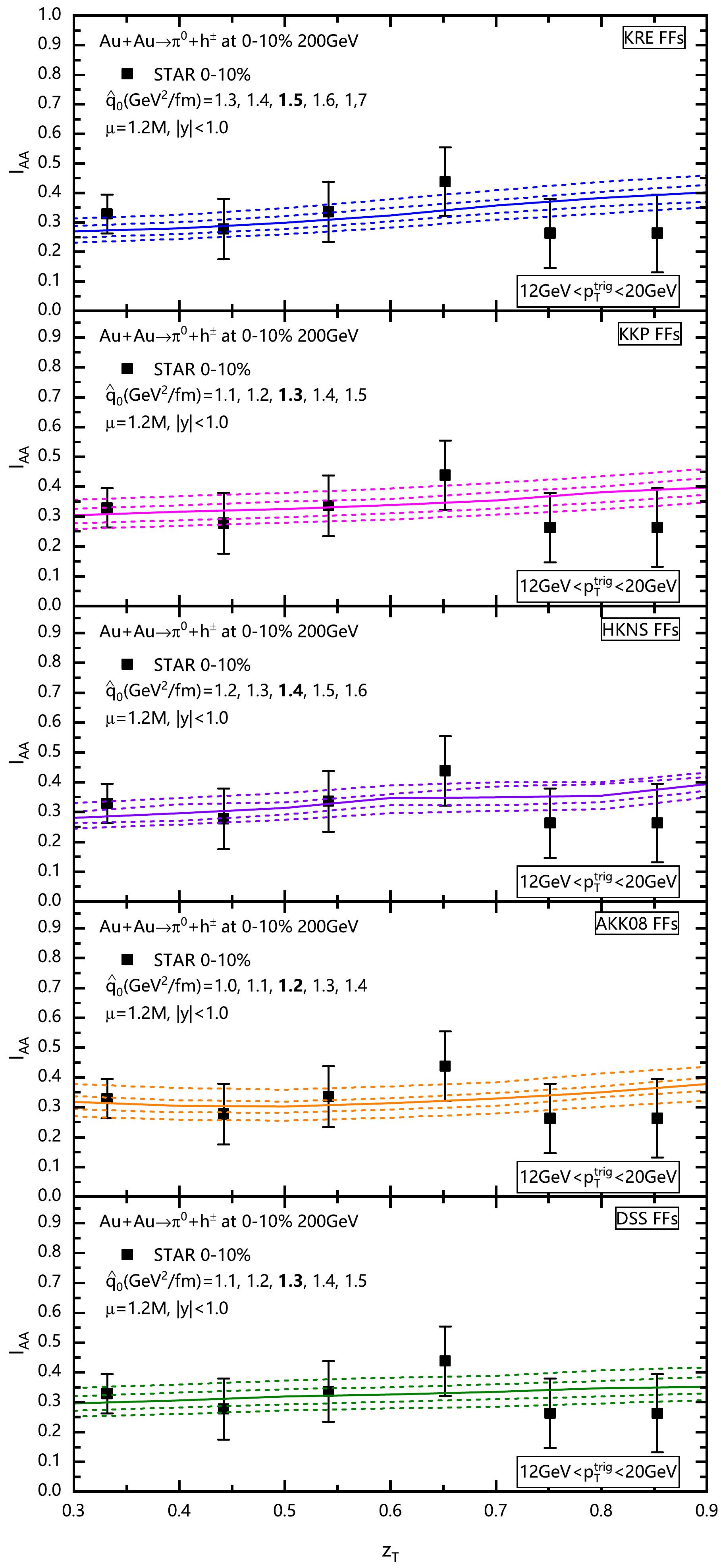}
\caption{The nuclear modification factor $R_{AA}$ for single hadron productions as a function of $p_{\rm T}$ (left panels) in $0-5\%$ $Au+Au$ collisions, and $I_{AA}$ for dihadron productions as a function of $z_{\rm T}$ (right panels) in $0-10\%$ $Au+Au$ collisions at $\sqrt{s_{\rm NN}}=200$ GeV. The experimental data are from Refs. \cite{Adare:2008qa, Adare:2012wg,STAR:2016jdz}. }
\label{fig:raa_rhic_s}
\end{figure}
\begin{figure}[htbp]
\centering
\includegraphics[width=0.32\textwidth]{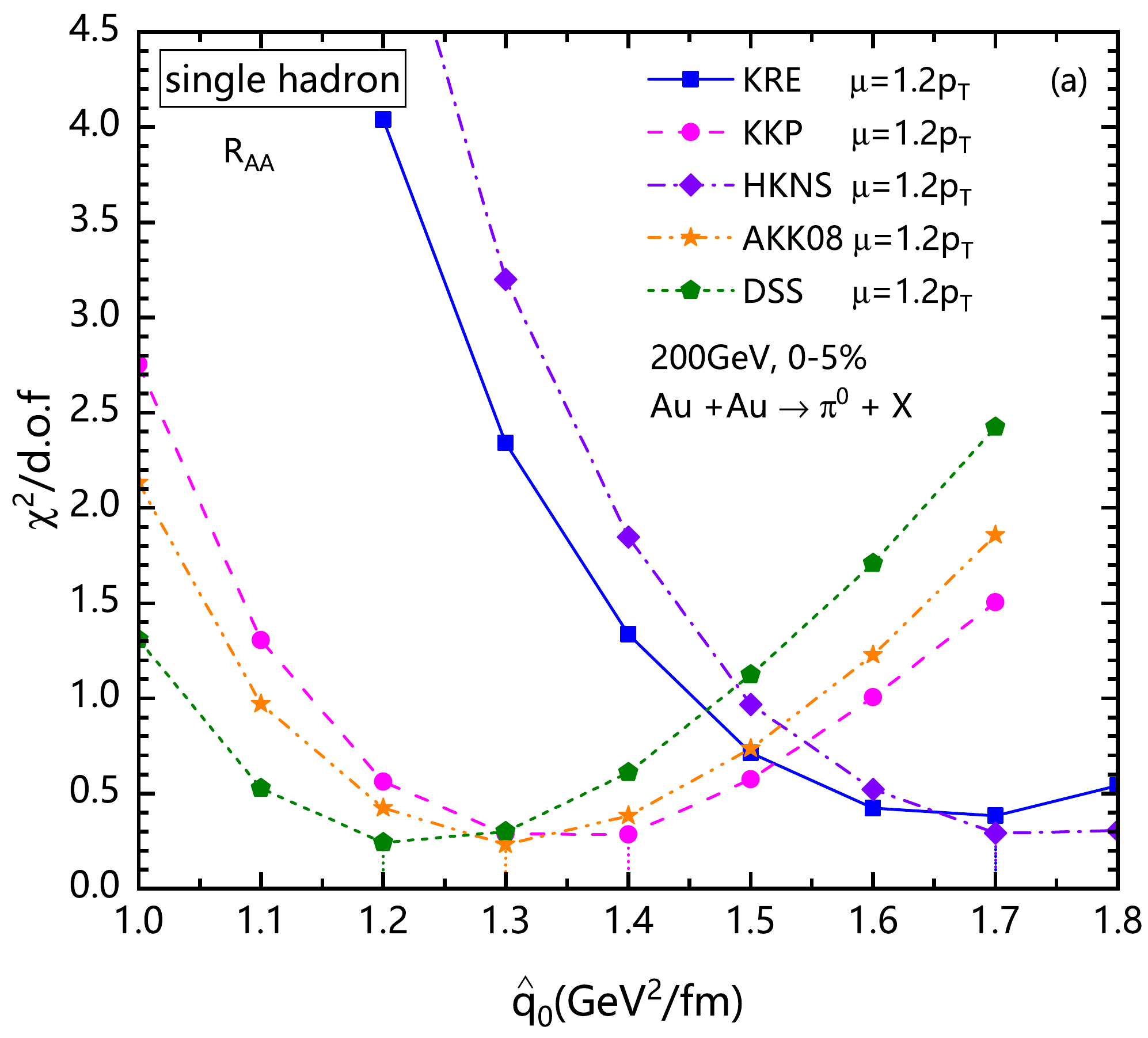}
\includegraphics[width=0.32\textwidth]{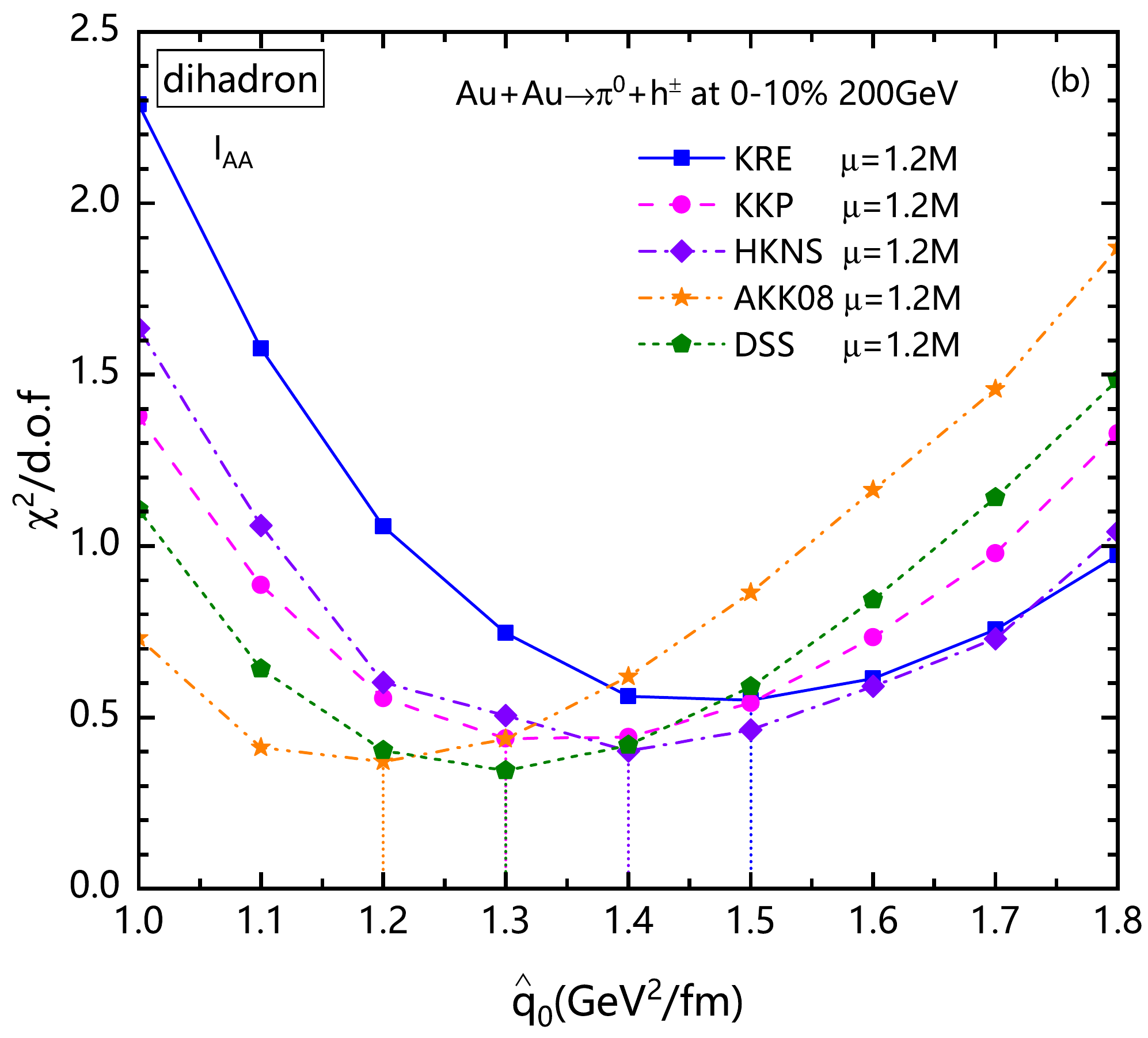}
\includegraphics[width=0.32\textwidth]{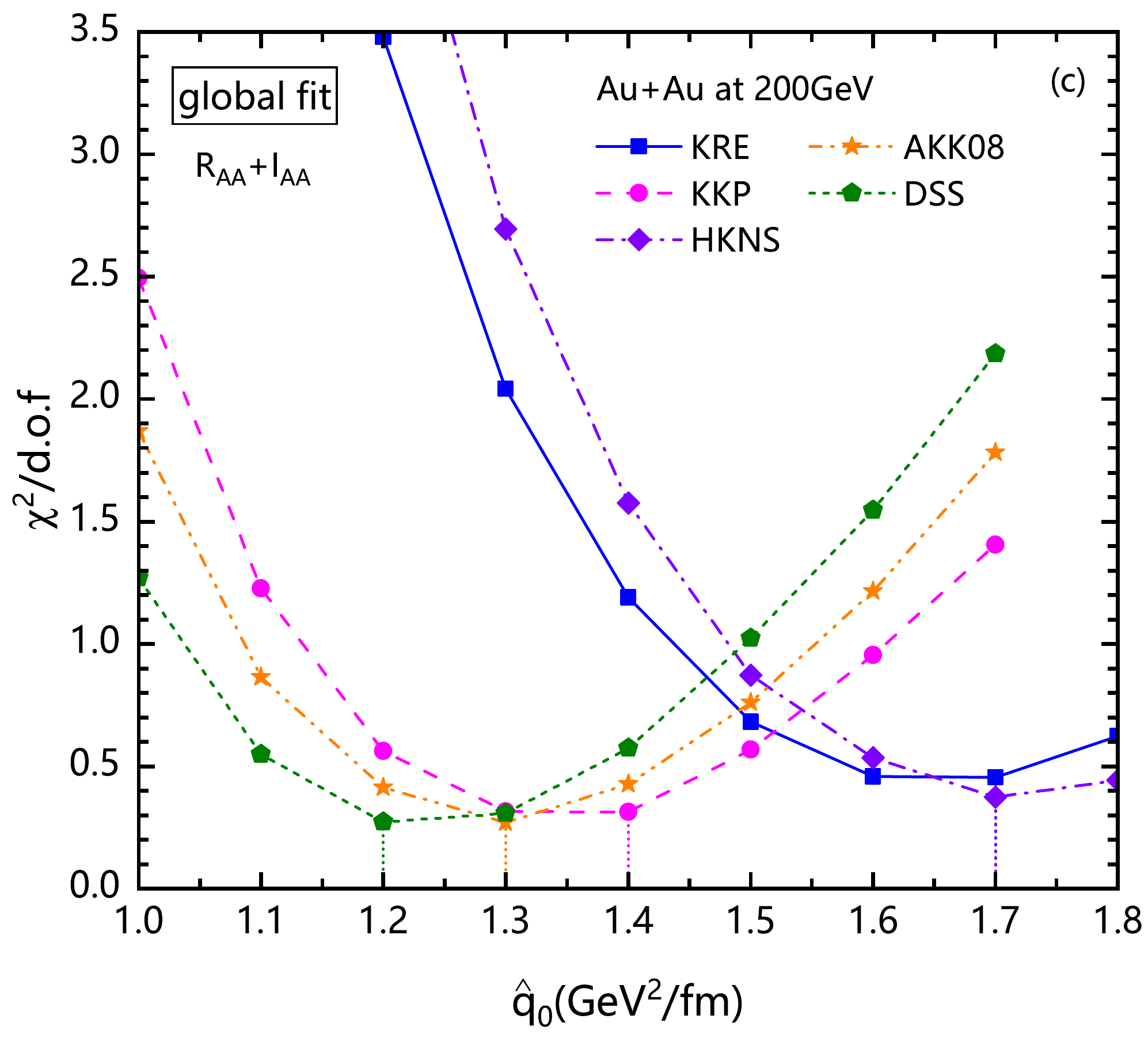}
\caption{The $\chi^2/d.o.f$ results from fitting to nuclear modification factors in central $Au+Au$ collisions at $\sqrt{s_{\rm NN}}=200$ GeV, (a) the fits to only single hadron $R_{AA}(p_{\rm T})$, (b) the fits to only dihadron $I_{AA}(z_{\rm T})$, and (c) the global fits to $R_{AA}(p_{\rm T})$ + $I_{AA}(z_{\rm T})$. 5 sets of fragmentation function parameterizations are used, respectively. }
\label{fig:kai2_rhic_s}
\end{figure}

\end{widetext}

The large differences in the values $\hat{q}_0=1.2-1.7$ GeV$^2$/fm are obtained by using different sets of FFs.
Considering parton-to-hadron contributions in Fig. \ref{fig:zD-z-pi_rhic_s} and \ref{fig:frac_rhic_s}, one can see that such differences for $\hat{q}_0$ extraction are mainly from and sensitive to the differences of gluon-to-hadron in FFs.
For example, as shown in Fig. \ref{fig:frac_rhic_s} with the different sets of FFs, one can get the least fraction of gluon-to-hadron contribution with KRE parametrization while the largest fraction with DSS parametrization. Since gluon energy loss is 9/4 times of quark energy loss, model calculations with KRE FFs give $\hat{q}_0=1.7$ GeV$^2$/fm larger than $\hat{q}_0=1.2$ GeV$^2$/fm with DSS FFs for the almost same total energy loss for $R_{AA}(p_{\rm T})$ to fit data well in central $A+A$ collisions.

\begin{widetext}
\begin{center}
\begin{figure}[htbp]
\includegraphics[width=0.60\textwidth]{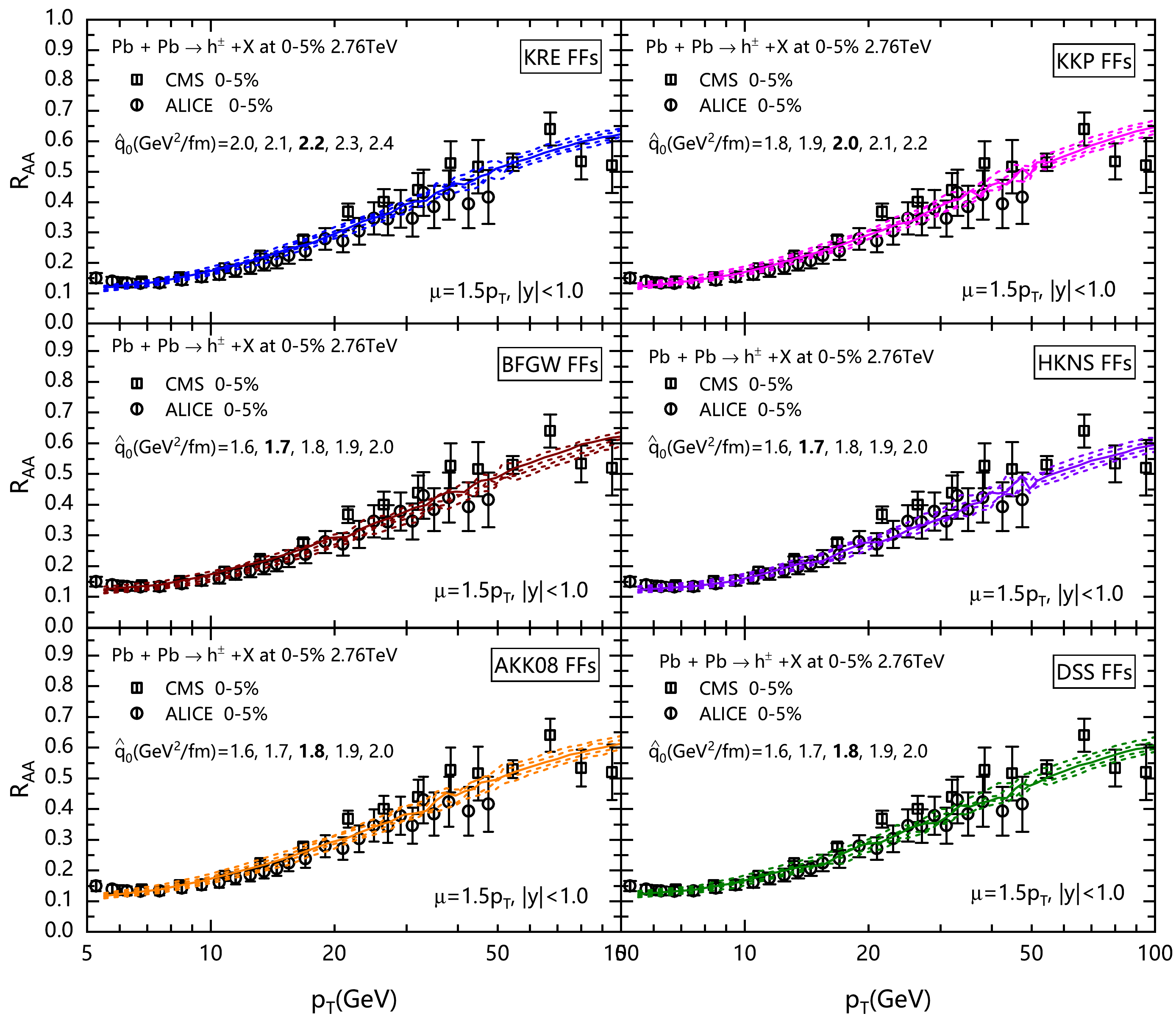}
\caption{The nuclear modification factor $R_{AA}$ for single hadron as a function of $p_{\rm T}$ compared to the experimental data from ALICE \cite{Abelev:2012hxa} and CMS \cite{CMS:2012aa} in $0-5\%$ $Pb+Pb$ collisions at $\sqrt{s_{\rm NN}}=2.76$ TeV.}
\label{fig:raa_lhc_s}
\end{figure}
\end{center}
\end{widetext}

\subsection{Extracting $\hat{q}_0$ in central $Pb+Pb$ collisions at $\sqrt{s_{\rm NN}}=2.76$ TeV}

Similar to at RHIC energy, we perfrom the same analyses for single hadron $R_{AA}(p_{\rm T})$ and dihadron $I_{AA}(p_{\rm T}^{\rm assoc})$ (corresponding to Eq. (\ref{eq:Iaa-pt})) in central $Pb+Pb$ collisions at $\sqrt{s_{\rm NN}}=2.76$ TeV.
BFGW fragmentation function parameterizations \cite{Bourhis:2000gs} is also applied for charged hadrons besides the other 5 sets of FFs.
Shown in Fig.~\ref{fig:raa_lhc_s} are the nuclear modification factor $R_{AA}(p_{\rm T})$ for single hadrons in 0-5\% $Pb+Pb$ collisions with the same scale $\mu = 1.5 p_{\rm T}$ for the 6 sets of FFs. In the meanwhile, the nuclear modification factor $I_{AA}(p_{\rm T}^{\rm assoc})$ for dihadrons with the same scale $\mu = 1.5 M$ are presented in Fig.~\ref{fig:iaa_alice_s} and Fig.~\ref{fig:iaa_cms_s}, which contain the results for different $p_{\rm T}^{\rm trig}$ ranges of the triggered hadrons, respectively. As shown in Fig.~\ref{fig:iaa_alice_s}, the trigger transverse momentum is chosen with a relatively small range ($8<p_{\rm T}^{\rm trig}<16$ GeV), and the theoretical results for $I_{AA}(p_{\rm T}^{\rm assoc})$ are compared to the ALICE experimental data \cite{ALICE:2011gpa}. The solid curve denotes the numerical result with the best fitting $\hat{q}_0$ obtained by the $\chi^2$ fitting to the experimental data.
In Fig.~\ref{fig:iaa_cms_s}, each panel with each set of FFs contains four sub-panels in which theoretical results are compared with the CMS data \cite{Conway:2013xaa} for $I_{AA}(p_{\rm T}^{\rm assoc})$ with four different large $p_{\rm T}^{\rm trig}$ ranges. The theoretical results fit data well with several appropriate values of $\hat{q}_0$ at the LHC energy.

\begin{widetext}
\begin{center}
\begin{figure}[htbp]
\includegraphics[width=0.60\textwidth]{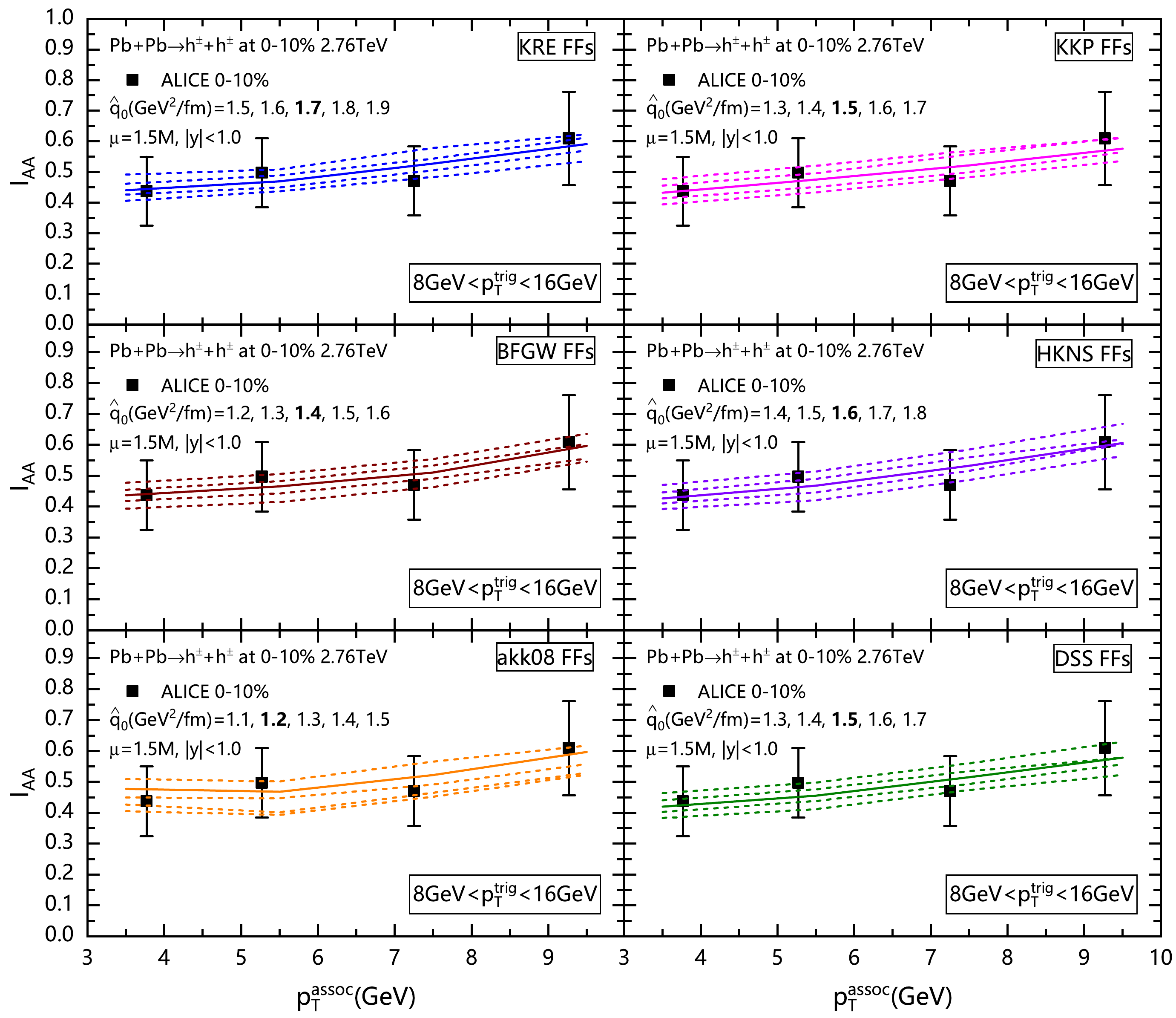}
\caption{The nuclear modification factor $I_{AA}$ as a function of $p_{\rm T}^{\rm assoc}$ for dihadrons compared to the experimental data from ALICE \cite{ALICE:2011gpa} in $0-10\%$ $Pb+Pb$ collisions at $\sqrt{s_{\rm NN}}=2.76$ TeV.}
\label{fig:iaa_alice_s}
\end{figure}
\end{center}
\end{widetext}

Shown in Fig.~\ref{fig:kai2_lhc_s} are the $\chi^2/d.o.f$ fits to nuclear modification factors in central $Pb+Pb$ collisions at $\sqrt{s_{\rm NN}}=2.76$ TeV, (a) the fits to only single hadron $R_{AA}(p_{\rm T})$, (b) the fits to only dihadron $I_{AA}(p_{\rm T}^{\rm assoc})$, and (c) the global fits to $R_{AA}(p_{\rm T})$ + $I_{AA}(p_{\rm T}^{\rm assoc})$. The fitted data include ALICE \cite{Abelev:2012hxa, ALICE:2011gpa} and CMS \cite{CMS:2012aa, Conway:2013xaa} data.
The 6 sets of fragmentation function parameterizations are used, respectively. $p_{\rm T} > 5$ GeV for single hadrons $R_{AA}(p_{\rm T})$ and $p_{\rm T}^{\rm assoc} > 3.5$ GeV for $h^{\pm}$-triggered away-side charged hadrons $I_{AA}(p_{\rm T}^{\rm assoc})$ are chosen to fit data. From the global-fit results of Fig. \ref{fig:kai2_lhc_s} (c), we can read the best-fitting values of $\hat{q}_0$ for each set of FFs as: $\hat{q}_0=2.2$ GeV$^2/$fm with KRE FFs, $\hat{q}_0=2.1$ GeV$^2/$fm with KKP FFs, $\hat{q}_0=1.8$ GeV$^2/$fm with BFGW FFs, $\hat{q}_0=1.7$ GeV$^2/$fm with HKNS FFs, $\hat{q}_0=1.9$ GeV$^2/$fm with AKK08 FFs, $\hat{q}_0=1.8$ GeV$^2/$fm with DSS FFs.
The difference in the extracted values of $\hat{q}_0=1.7 - 2.2$ GeV$^2/$fm from different sets of FFs is caused by the different contributions of gluon-to-hadron FFs, as shown in Fig. \ref{fig:zD-z-h_lhc_s} and \ref{fig:frac_lhc_s}, similarly to the case of RHIC energy.

\begin{widetext}
\begin{center}
\begin{figure}[htbp]
\subfigure{
\includegraphics[width=0.32\textwidth]{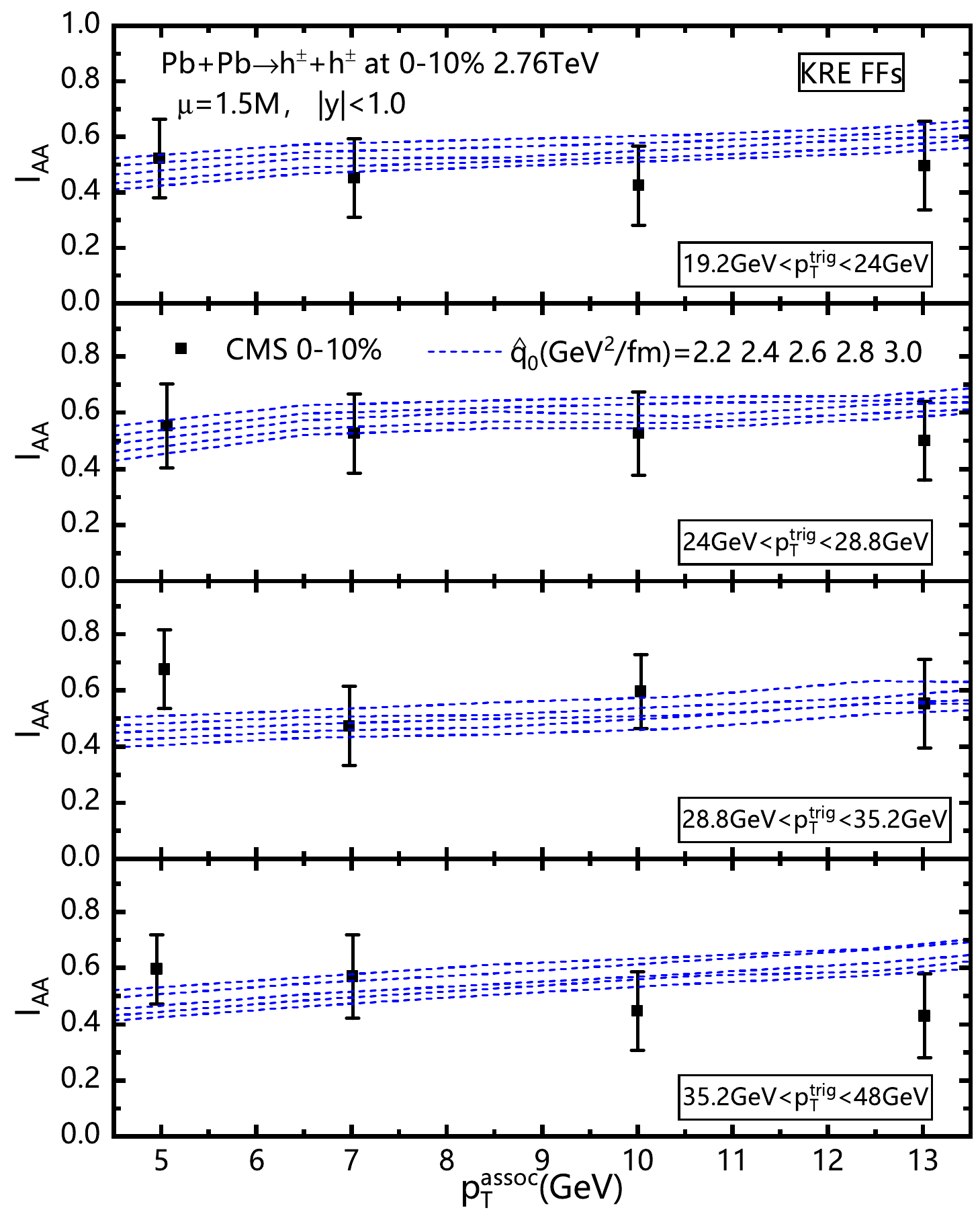}
\includegraphics[width=0.32\textwidth]{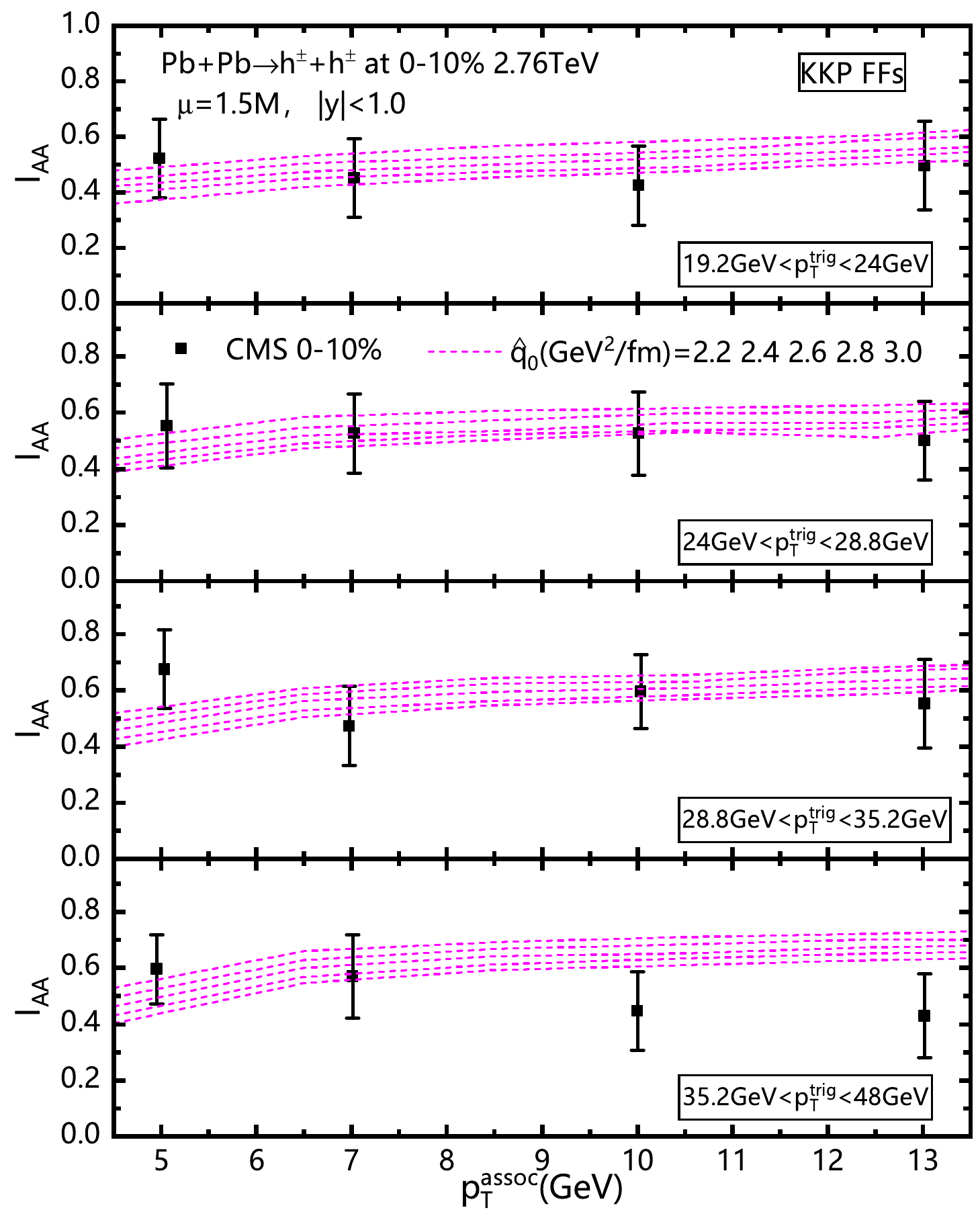}
\includegraphics[width=0.32\textwidth]{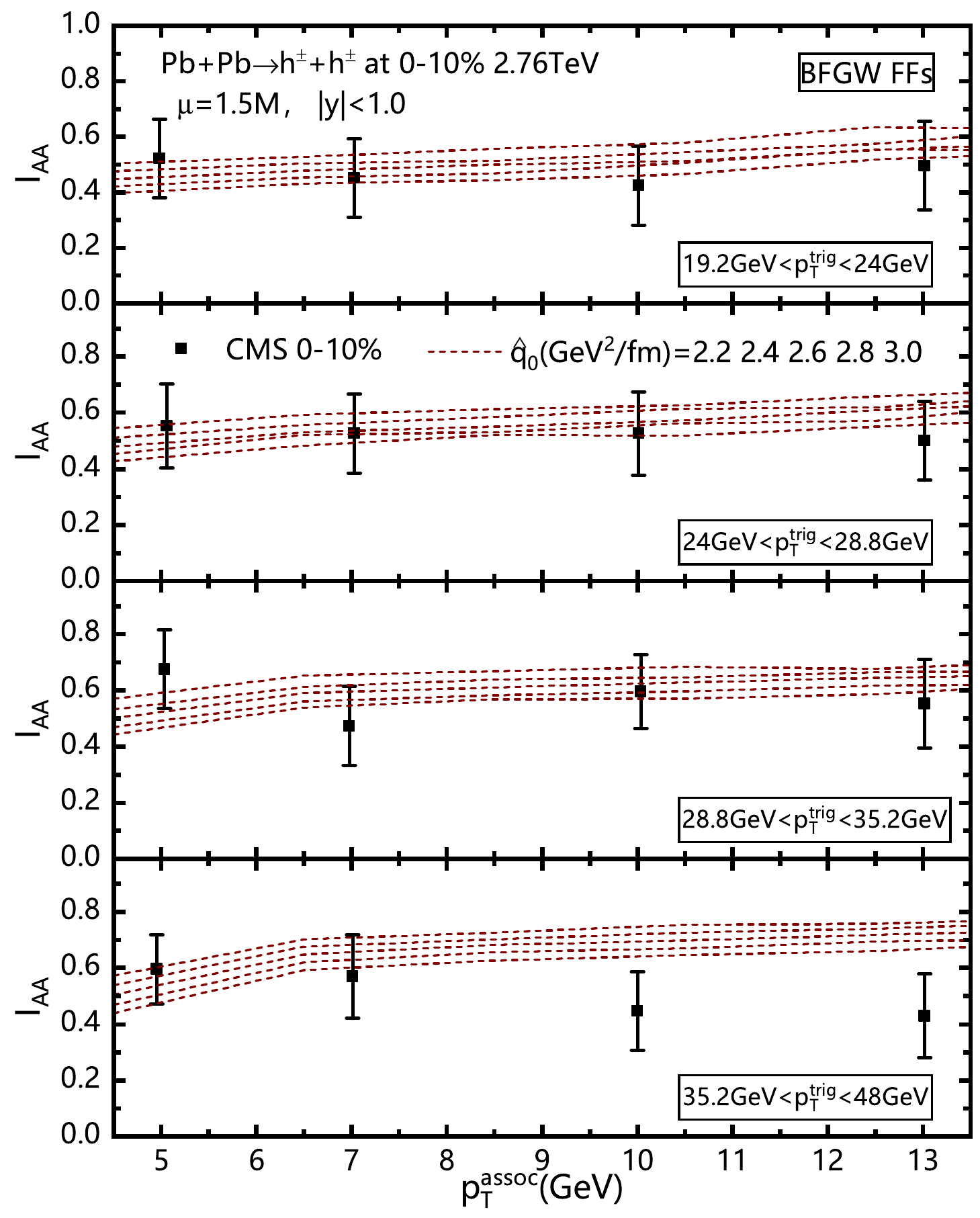}
}
\quad
\subfigure{
\includegraphics[width=0.32\textwidth]{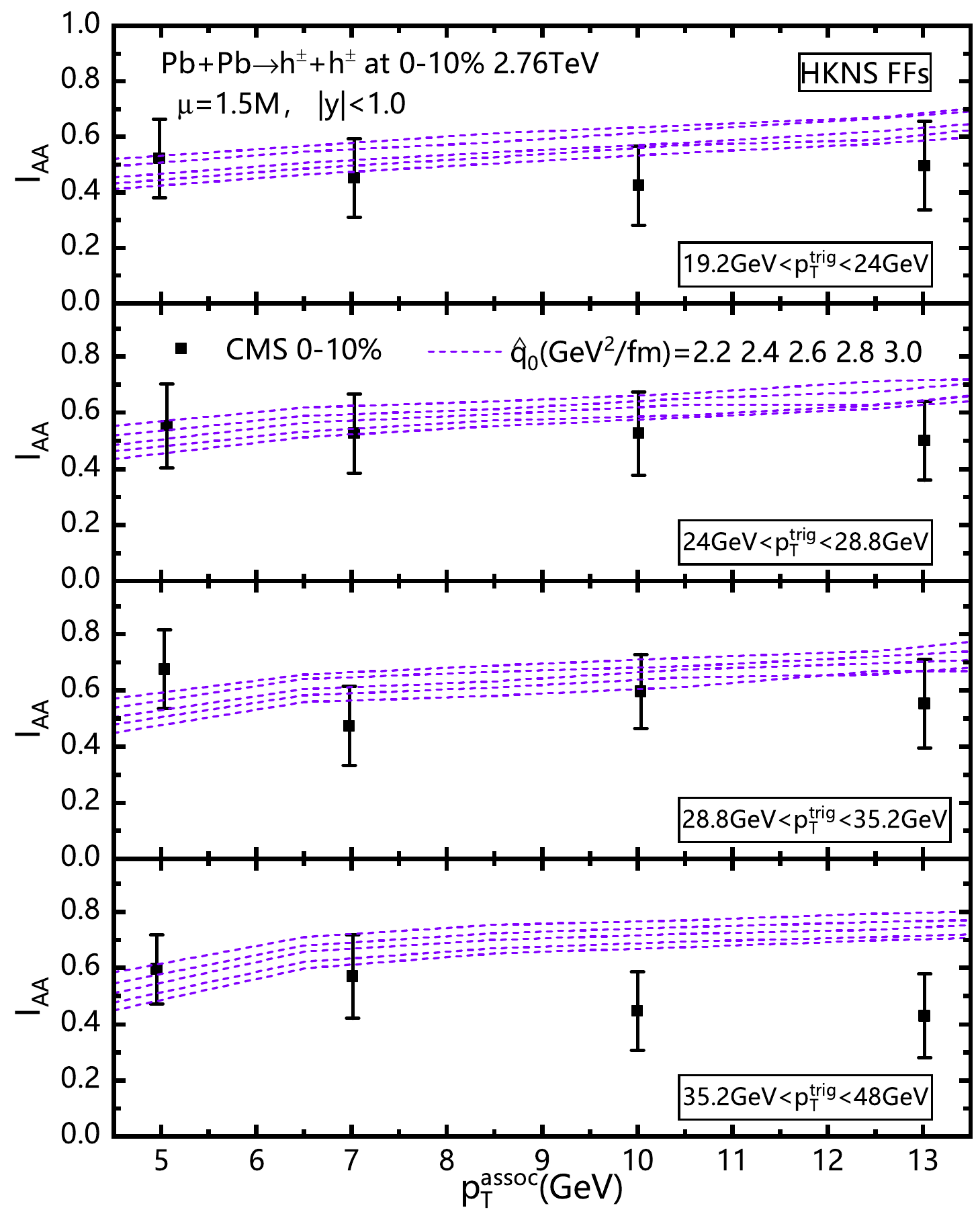}
\includegraphics[width=0.32\textwidth]{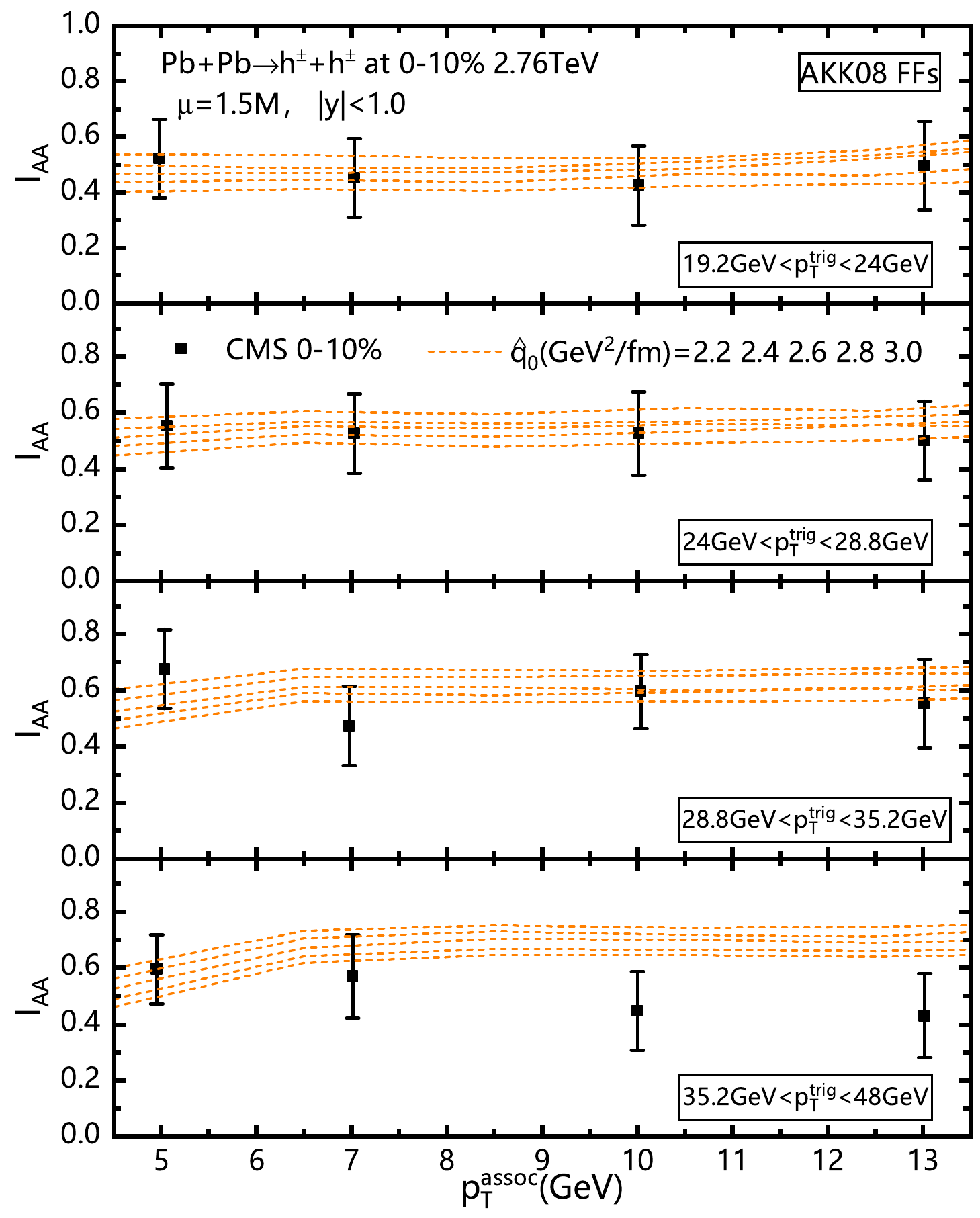}
\includegraphics[width=0.32\textwidth]{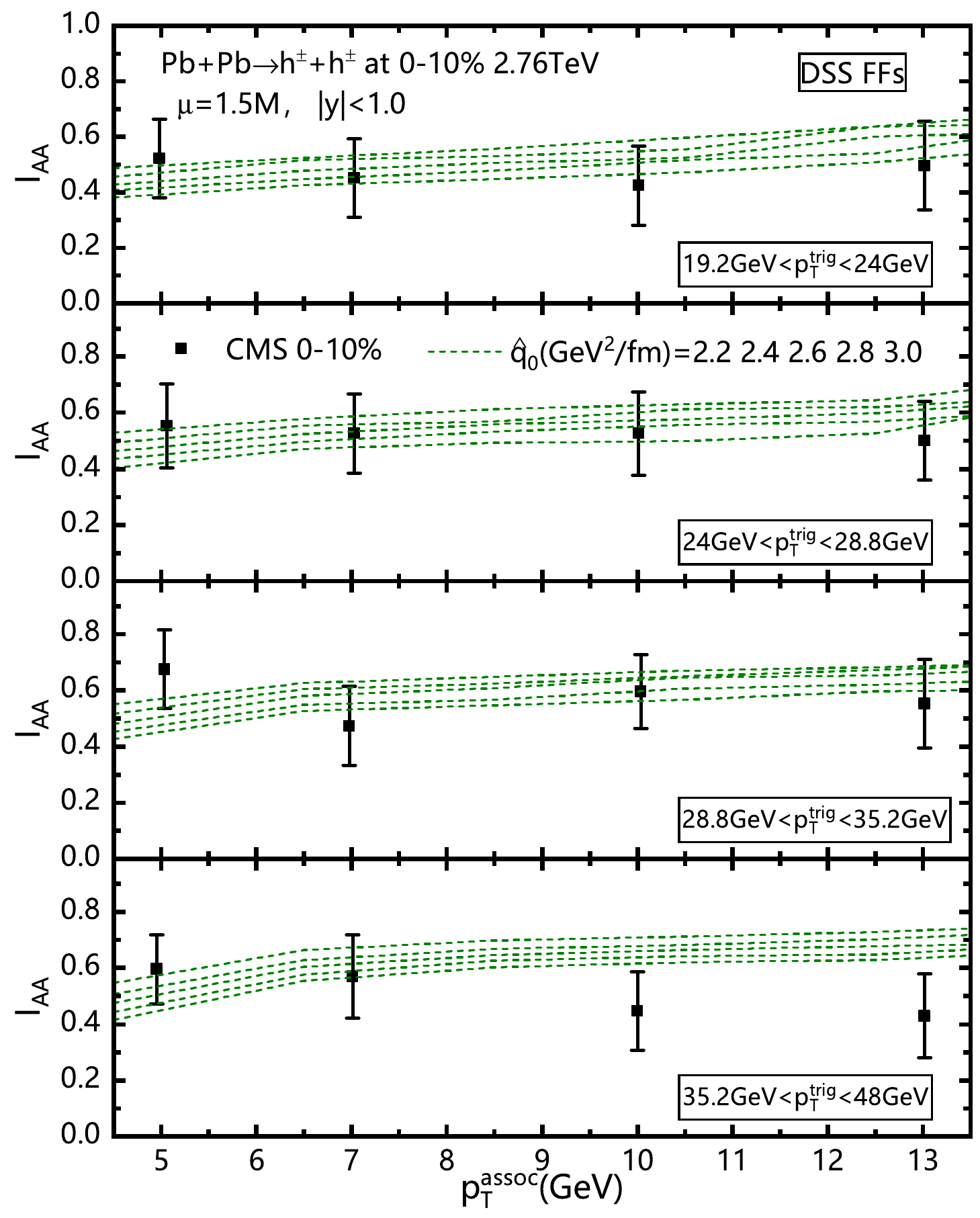}
}
\caption{The nuclear modification factor $I_{AA}$ as a funtion of $p_{\rm T}^{\rm assoc}$ for dihadrons with different $p_{\rm T}^{\rm trig}$ ranges compared to the experimental data from CMS \cite{Conway:2013xaa} in $0-10\%$ $Pb+Pb$ collisions at $\sqrt{s_{\rm NN}}=2.76$TeV.}
\label{fig:iaa_cms_s}
\end{figure}
\end{center}
\begin{center}
\begin{figure}[htbp]
\includegraphics[width=0.32\textwidth]{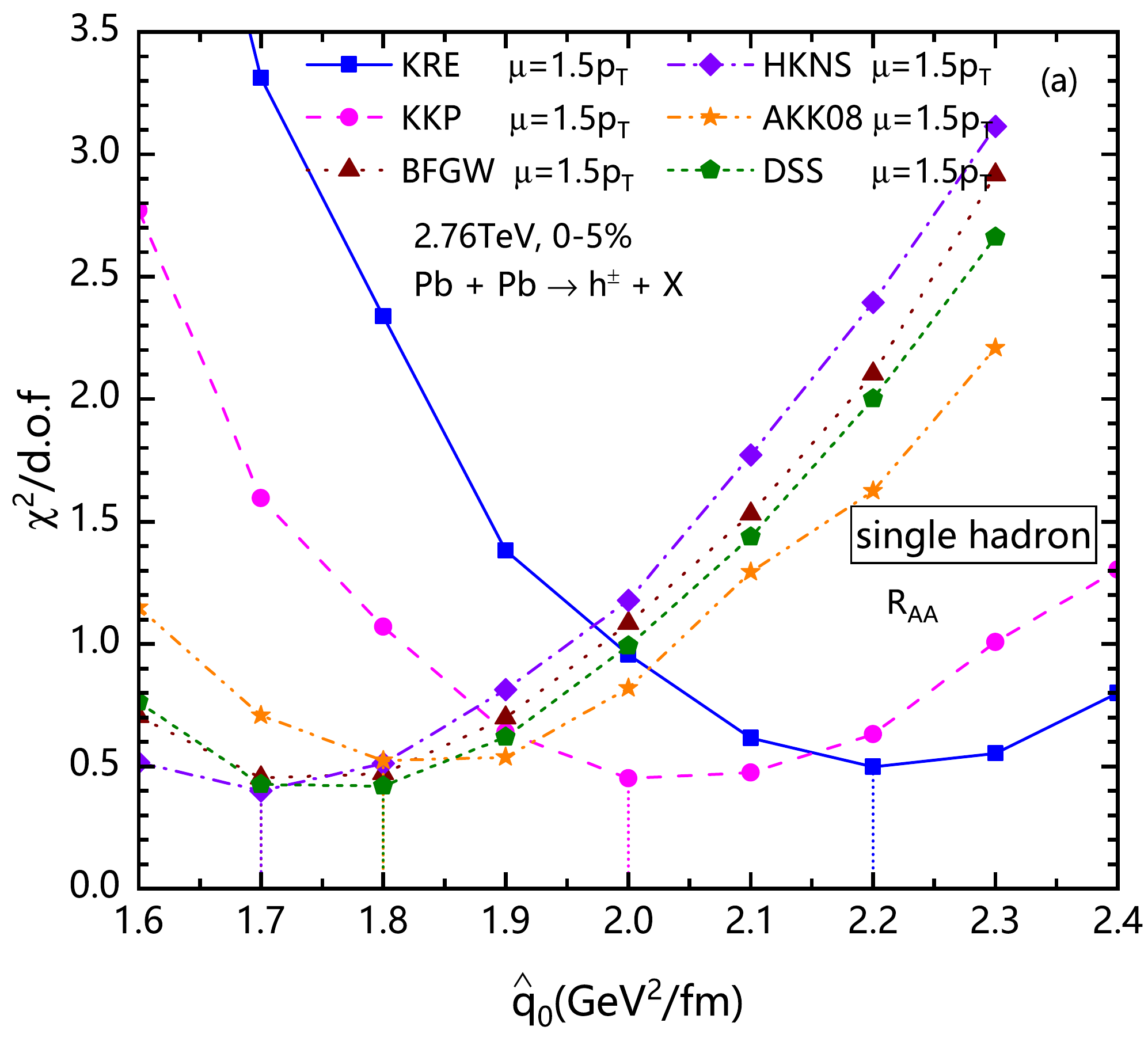}
\includegraphics[width=0.32\textwidth]{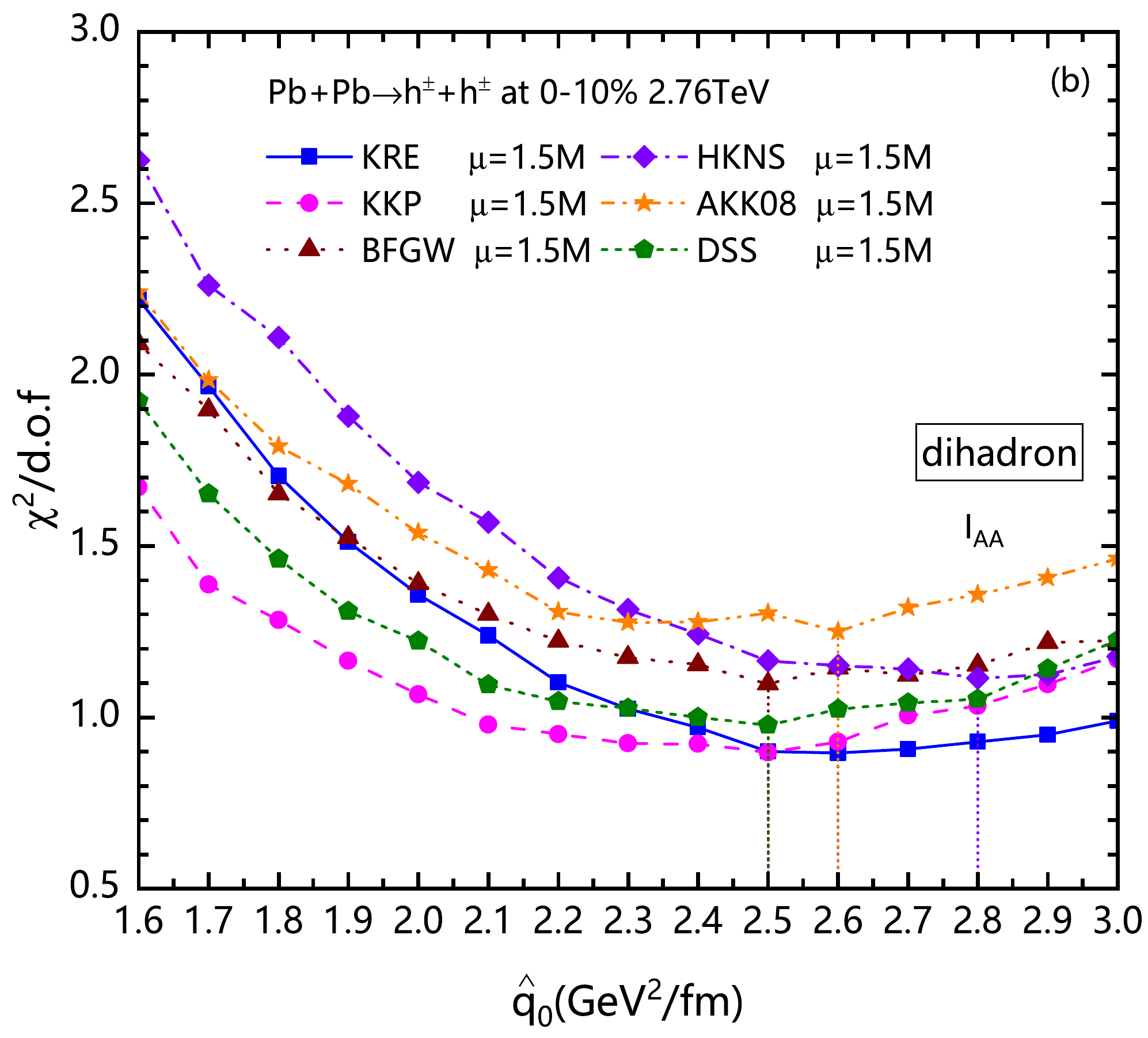}
\includegraphics[width=0.32\textwidth]{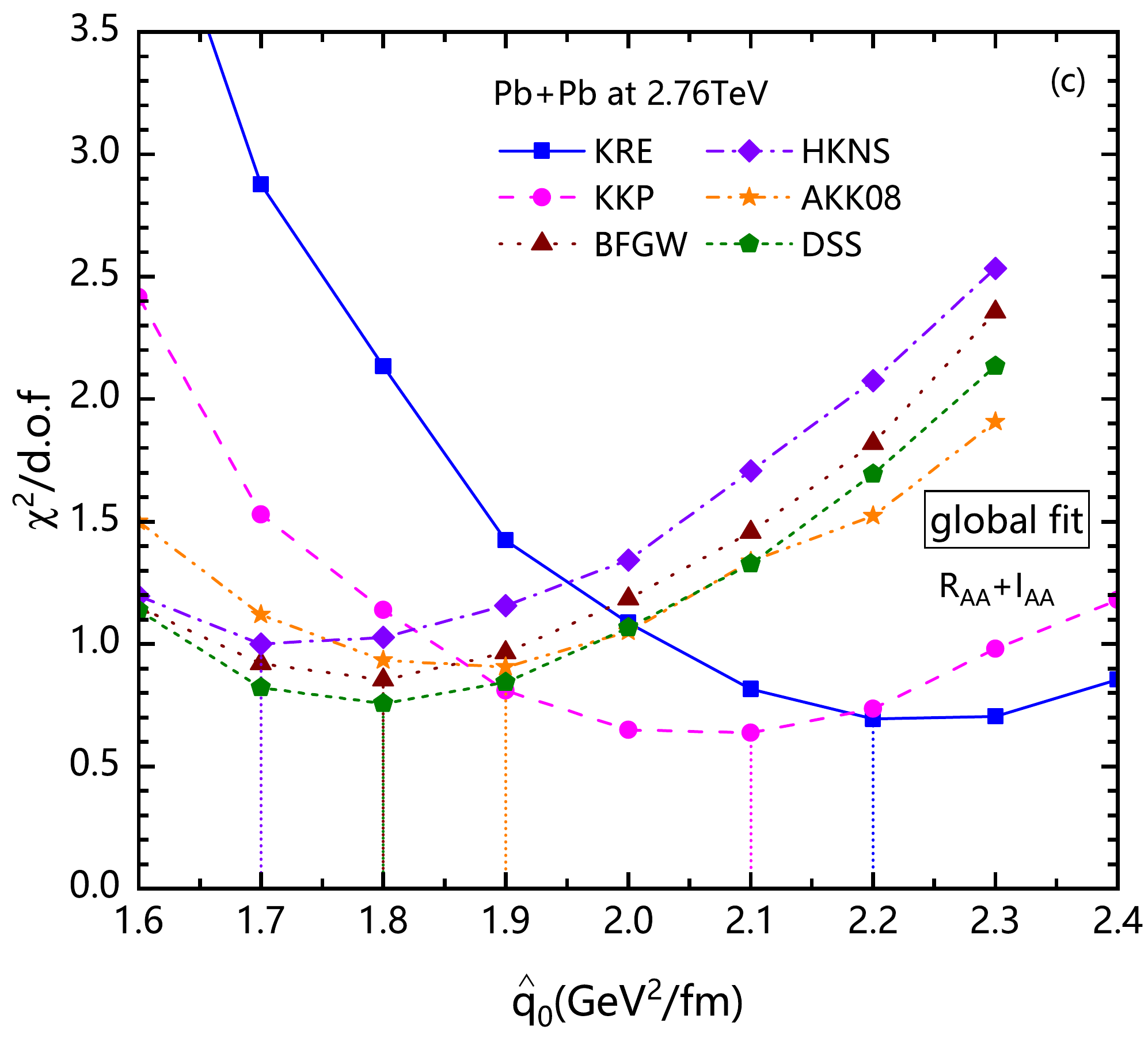}
\caption{The $\chi^2/d.o.f$ results from fitting to nuclear modification factors in central $Pb+Pb$ collisions at $\sqrt{s_{\rm NN}}=2.76$ TeV, (a) the fits to only single hadron $R_{AA}(p_{\rm T})$ \cite{Abelev:2012hxa, CMS:2012aa}, (b) the fits to only dihadron $I_{AA}(p_{\rm T}^{\rm assoc})$ \cite{ALICE:2011gpa, Conway:2013xaa} data, and (c) the global fits to $R_{AA}(p_{\rm T})$ + $I_{AA}(p_{\rm T}^{\rm assoc})$. The 6 sets of fragmentation function parameterizations are used, respectively.}
\label{fig:kai2_lhc_s}
\end{figure}
\end{center}

\end{widetext}

Note that in panels (a) and (b) of Fig.~\ref{fig:kai2_lhc_s}, with the same set of FFs, the $\hat{q}_0$ value from dihadron suppressions is larger than that from single hadron suppressions, which is mainly due to the different production positions of the initial parton jets for single hadrons and dihadrons.
On one hand, the single hadrons are mainly contributed from the surface emissions of single jets perpendicular to the surface of the hot medium, which means that a large number of single jets produced at the center of the medium are melted by the medium \cite{Zhang:2007ja}. In contrast, the dihadrons are contributed from a combination of tangential dijets generated at the medium surface and punching-through dijets created at the matter center with limited energy loss \cite{Zhang:2007ja}. Thus in an $A+A$ event the average jet energy loss for dihadrons is smaller than single hadrons.
If one wants to obtain similar hadron suppressions shown as the experimental measurements, larger average energy loss is needed for dihadrons than single hadrons.
On the other hand, the fraction of the punching-through jets becomes larger with the increasing $p_{\rm T}^{\rm trig}$ \cite{Zhang:2008fh}. That means as the $p_{\rm T}^{\rm trig}$ increases, the suppression of dihadrons should become less, which is not manifested obviously by the experimental data with relatively considerable uncertainty \cite{Conway:2013xaa}.
With the larger $p_{\rm T}^{\rm trig}$, one need to adjust $\hat{q}_0$ to be larger for $I_{AA}$ to fit the experimental measurements.
Therefore, in the global fit for $I_{AA}$ with all $p_{\rm T}^{\rm trig}$ ranges, the values of $\hat{q}_0$ extracted from dihadrons will be enlarged.
Such significant difference of $\hat{q}_0$ between single hadrons and dihadrons can be attenuated by the abundant and accurate single hadron data, as shown in panel (c) of Fig.~\ref{fig:kai2_lhc_s}.

\subsection{The uncertainty for $\hat{q}_0$ extraction due to FFs}

With the same scale in the NLO pQCD parton model including different sets of FFs for large $p_{\rm T}$ hadron productions in central $A+A$ collisions at RHIC and the LHC, we calculate both the nuclear modification factors for single hadron and dihadron suppressions due to jet quenching, and make a global fit to data to extract the jet energy loss parameter $\hat{q}_0$, as shown in Fig.~\ref{fig:kai2_rhic_s} (c) and Fig.~\ref{fig:kai2_lhc_s} (c).
To illustrate clearly the uncertainties from different FFs, we summarize all the extracted values for $\hat{q}_0$ to show in the left panel of Fig.~\ref{fig:qhat_uncertainty}.
The blue squares are for the scaled jet transport coefficient of $\hat{q}_0/T_0^3$ in central $Au+Au$ collisions at $\sqrt{s_{\rm NN}}=200$ GeV, and the center temperature of the medium in the initial time is chosen as $T_0=370$ MeV.
The pink circles denote the results for central $Pb+Pb$ collisions at $\sqrt{s_{\rm NN}}=2.76$ TeV, and $T_0=480$ MeV.
The dotted-curve boxes indicate the uncertainties for $\hat{q}_0/T_0^3$ from the different sets of FFs, and we can read that with the same scales in the model, $\hat{q}/T^3 = 4.74 - 6.72$ at $T = 370$ MeV and $\hat{q}/T^3 = 3.07 - 3.98$ at $T = 480$ MeV.

\begin{widetext}

\begin{center}
\begin{figure}[htbp]
\includegraphics[width=0.4\textwidth]{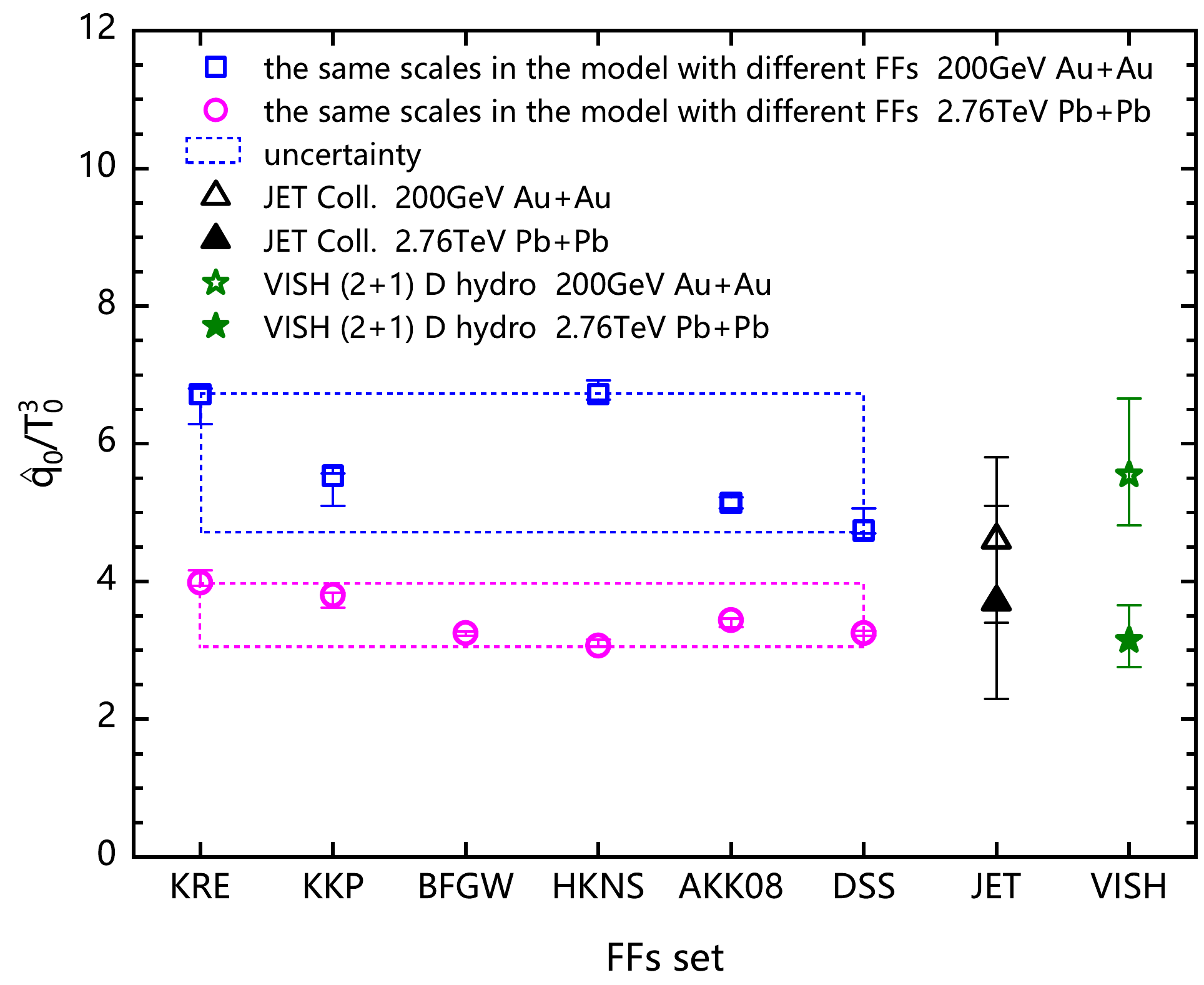}
\includegraphics[width=0.4\textwidth]{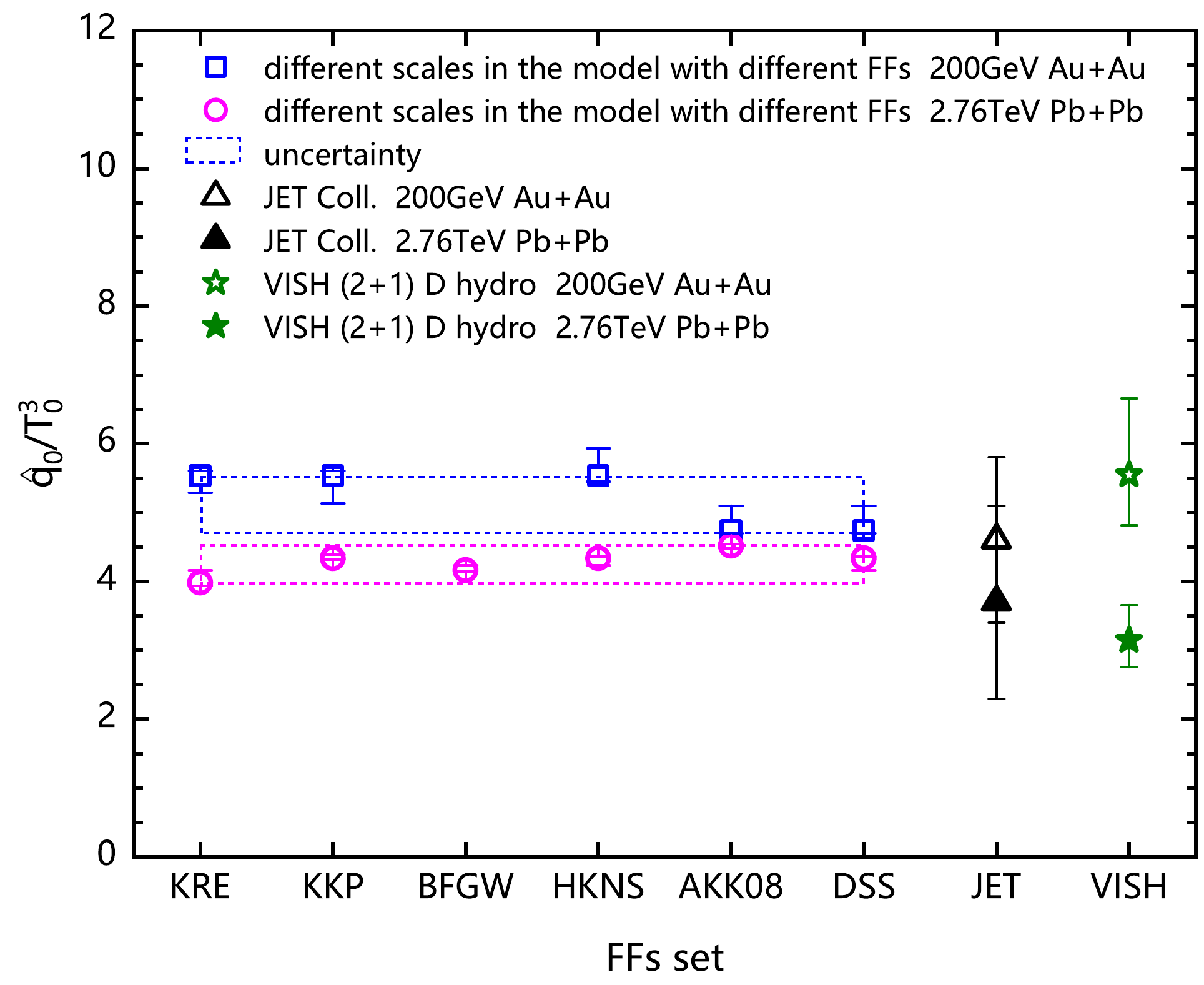}
\caption{Comparisons between the best-fitting values for the scaled $\hat{q}_0/T_0^3$ extracted with different sets of FFs in the theoretical model at RHIC and the LHC, left panel for the case of the same scales in the model with different sets of FFs, and right panel for the different scales. These results are also compared to those from JET collaboration \cite{Burke:2013yra} and a previous theoretical study \cite{Xie:2020zdb} including the VISH (2+1)-dimension hydrodynamics \cite{Song:2007ux,Qiu:2011hf}.}
\label{fig:qhat_uncertainty}
\end{figure}
\end{center}

\end{widetext}

In contrast to the case of the same scale $\mu$ in the theoretical model for different sets of FFs, we also accomplish the $\hat{q}_0$ extraction with different scales in the model for different sets of FFs, as shown in the right panel of Fig. \ref{fig:qhat_uncertainty}. The suitable scale is included in the model for each set of FFs to make numerical hadron spectra fit data well in $p+p$ collisions.
The best fitting values of $\hat{q}_0/T_0^3$ with different scales are summarized as, $\hat{q}/T^3 = 4.74 - 5.53$ at $T = 370$ MeV in central $Au+Au$ collisions at $\sqrt{s_{\rm NN}}=200$ GeV, and $\hat{q}/T^3 = 3.98 - 4.52$ at $T = 480$ MeV in central $Pb+Pb$ collisions at $\sqrt{s_{\rm NN}}=2.76$ TeV, respectively.

The $\hat{q}_0/T_0^3$ uncertainty from different sets of FFs is reduced significantly when the hadron-spectrum baseline of $p+p$ collisions is well adjusted with different scales. For a given set of FFs used in the NLO pQCD parton model for large $p_{\rm T}$ hadron spectra in $p+p$ collisions, the ratio of gluon over quark contributions to hadrons decreases with the increasing of the chosen scale, as shown in Fig.~\ref{fig:frac_rhic_s} and \ref{fig:frac_lhc_s}, or Fig. \ref{fig:frac_akk08_rhic} and \ref{fig:frac_akk08_lhc} of Appendix B. The extracted $\hat{q}_0$ in central $A+A$ collisions increases with the increasing of the chosen scale due to gluon energy loss being $9/4$ times of quark energy loss, so the $\hat{q}_0/T_0^3$ uncertainty from different sets of FFs is balanced to some extent relative to the case of the same scale for different sets of FFs. The more detailed discussions are included in the following Appendix.

Finally, we also compare our results to those from JET collaboration \cite{Burke:2013yra} (black triangles) and a previous theoretical study \cite{Xie:2020zdb} (green stars) including the VISH (2+1)-dimension hydrodynamics \cite{Song:2007ux,Qiu:2011hf}. All the results are consistent, although different theoretical models or hydrodynamics cause some systematic uncertainty for the accurate extraction of jet transport coefficient.

\section{SUMMARY} \label{sec:summary}

Based on the next-to-leading-order perturbative QCD parton model incorporating modified fragmentation functions, we have studied single hadron and dihadron productions in high-energy heavy-ion collisions at both RHIC and the LHC energies.
The fragmentation functions are modified due to jet quenching in central $A+A$ collisions, the strength of which is characterized by the jet transport coefficient $\hat q$. The six current sets of NLO fragmentation function parameterizations are used in actual calculations for the nuclear modification factors $R_{AA}$ for single hadrons and $I_{AA}$ for dihadrons. We perform a global $\chi^2$ fitting to both $R_{AA}$ and $I_{AA}$ data to extract the jet transport coefficient in the initial time at the center of the created QGP medium, and obtain $\hat{q}/T^3 = 4.74 - 6.72$ at $T = 370$ MeV in central $Au+Au$ collisions at $\sqrt{s_{\rm NN}}=200$ GeV, and $\hat{q}/T^3 = 3.07 - 3.98$ at $T = 480$ MeV in central $Pb+Pb$ collisions at $\sqrt{s_{\rm NN}}=2.76$ TeV.

The numerical results show that the significant uncertainties for $\hat{q}/T^3$ extraction are mainly brought by the different contributions of gluon-to-hadron in the different sets of fragmentation function parameterizations due to gluon energy loss being $9/4$ times of quark energy loss. The uncertainties are reduced, if the suitable scale $\mu$ is chosen in the NLO pQCD parton model with each set of current fragmentation function parameterizations to fit data well for large $p_{\rm T}$ hadron spectra in $p+p$ collisions. However, the accurate parameterizations from a forthcoming global fit of parton-to-hadron fragmentation functions \cite{dEnterria:2013sgr} will help to constrain the uncertainties for jet quenching parameter extractions.

\section*{ACKNOWLEDGMENTS}
 This work is supported in by National Natural Science Foundation of China under Grants No. 11935007, Guangdong Major Project of Basic and Applied Basic Research No. 2020B0301030008, and Science and Technology Program of Guangzhou No. 2019050001.

\section*{APPENDIX} \label{sec:appendix}
\subsection{The $\hat{q}_0$ extractions with different scales for different sets of FFs}

\begin{figure}[htbp]
\includegraphics[width=0.38\textwidth]{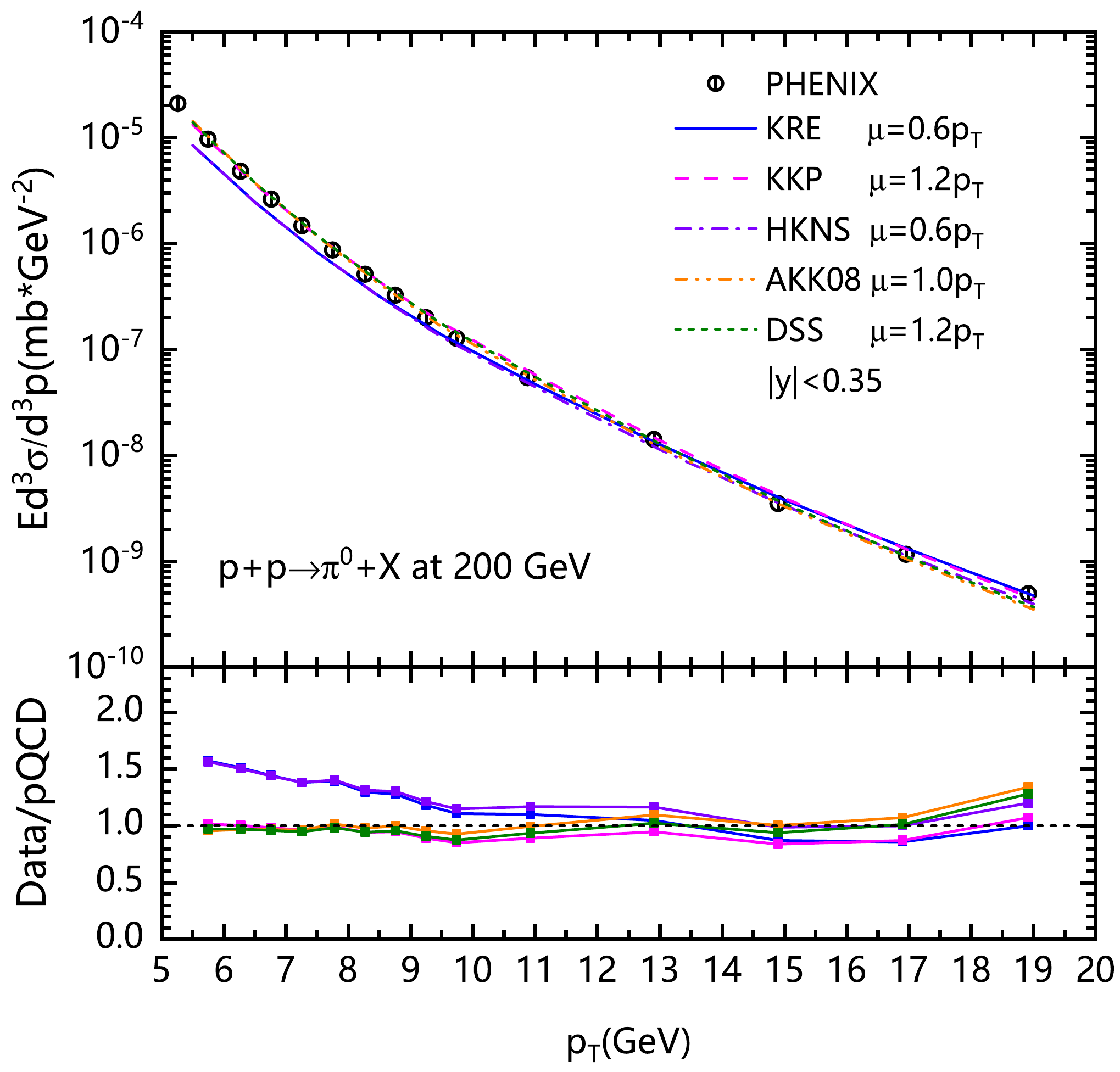}
\caption{$\pi^0$ spectra given by the NLO pQCD parton model with the befitted scale $\mu$ for each set of FFs in $p+p$ collisions at $\sqrt{s_{\rm NN}}=200$ GeV (upper panel), and spectrum ratios of data over the theoretical results (lower panel). The data are from Ref. \cite{Adare:2008qa}. }
\label{fig:pp_rhic_d}
\end{figure}

To check the nuclear or medium effects in $A+A$ collisions, phenomenologically, one need a suitable baseline of $p+p$ collisions by adjusting the scale $\mu$ to fit data well in $p+p$ collisions. Shown in the upper panel of Fig.~\ref{fig:pp_rhic_d} are the numerical results for $\pi^0$ hadron spectra at large transverse momentum $p_{\rm T}$ with the 5 sets of FFs in $p+p$ collisions at $\sqrt{s_{\rm NN}}=200$ GeV. Here the suitable scale $\mu$ is chosen in the theoretical model with each set of FFs for fitting to the experimental data \cite{Adare:2008qa}, respectively. The lower panel of Fig.~\ref{fig:pp_rhic_d} are for the ratios of the experimental data over theoretical calculations. One can see that with the appropriate scales $\mu$ in the model with different sets of FFs, the numerical results fit the data better relative to those in Fig.~\ref{fig:pp_rhic_s}.

\begin{figure}[htbp]
\includegraphics[width=0.38\textwidth]{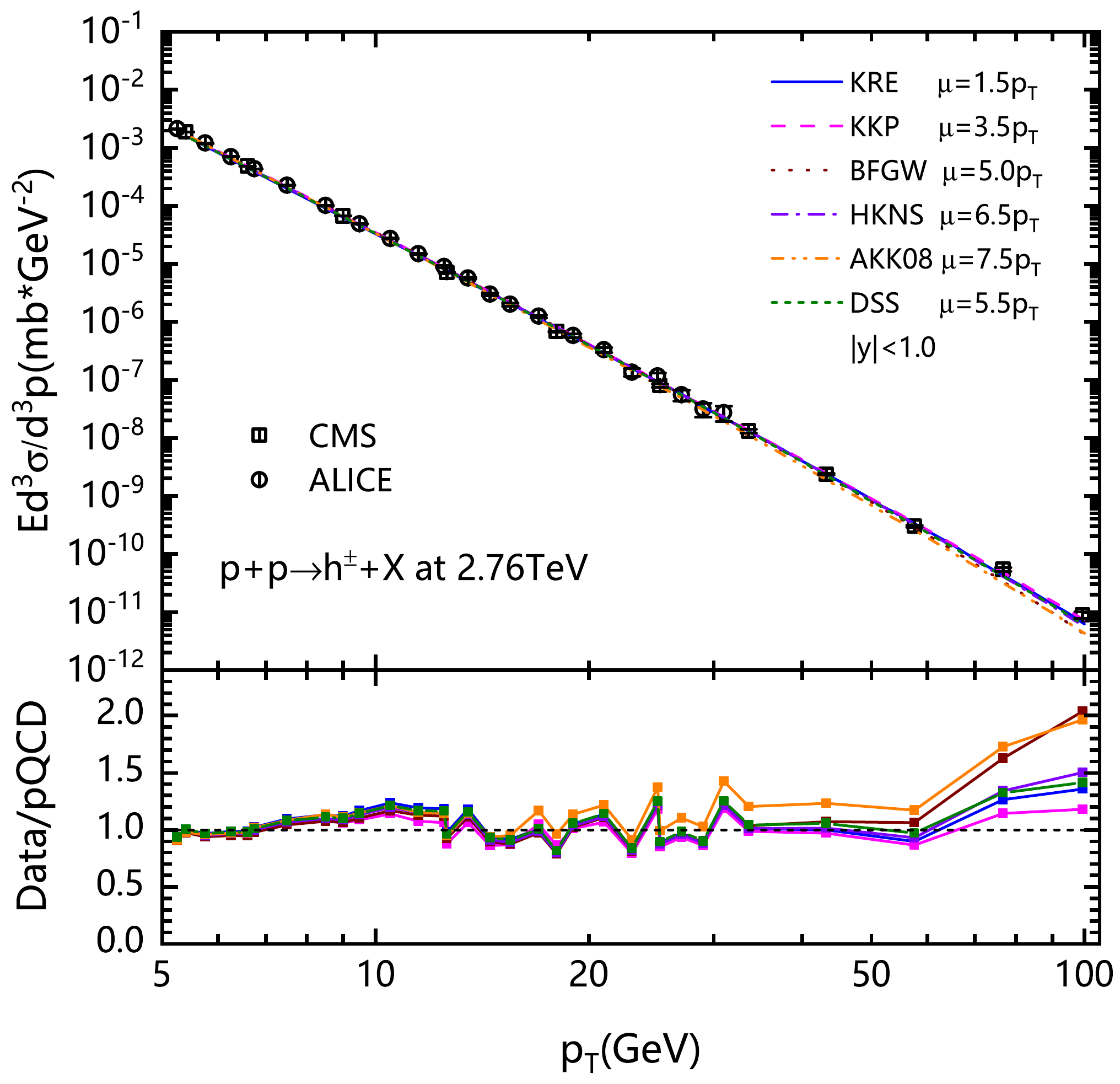}
\caption{Charged hadron spectra given by the NLO pQCD parton model with the befitted scale $\mu$ in $p+p$ collisions at $\sqrt{s_{\rm NN}}=2.76$ TeV (upper panel), and spectrum ratios of data over the theoretical results (lower panel). The data are from Refs. \cite{CMS:2012aa, Abelev:2012hxa}.}
\label{fig:pp_lhc_d}
\end{figure}

Similarly, in $p+p$ collisions at $\sqrt{s_{\rm NN}}=2.76$ TeV the charged hadron spectra are also calculated with different scales in the NLO pQCD parton model for the 6 sets of FFs, as shown in the upper panel of Fig.~\ref{fig:pp_lhc_d}. The ratios of the experimental data over theoretical calculations are shown in the lower panel. With the appropriate scales, the theoretical results fit data very well.

Using the above hadron spectra in $p+p$ collisions as baselines, we extract the jet quenching parameter from single hadron and dihadron suppressions with suitable scale in each set of FFs. Shown in Fig.~\ref{fig:kai2_rhic_d} are the $\chi^2/d.o.f$ fits to nuclear modification factors in central $Au+Au$ collisions at $\sqrt{s_{\rm NN}}=200$ GeV: panel (a) the fits to only single hadron $R_{AA}(p_{\rm T})$, panel (b) the fits to only dihadron $I_{AA}(z_{\rm T})$, and panel (c) the global fits to $R_{AA}(p_{\rm T})$ + $I_{AA}(z_{\rm T})$. From the panel (c), we can read the best-fitting values of jet transport coefficient: $\hat{q}_0=1.4$ GeV$^2$/fm with KRE FFs at $\mu=0.6p_{\rm T}$, $\hat{q}_0=1.4$ GeV$^2$/fm with KKP FFs at $\mu=1.2p_{\rm T}$, $\hat{q}_0=1.4$ GeV$^2$/fm with HKNS FFs at $\mu=0.6p_{\rm T}$, $\hat{q}_0=1.2$ GeV$^2$/fm with AKK08 FFs at $\mu=1.0p_{\rm T}$, and $\hat{q}_0=1.2$ GeV$^2$/fm with DSS FFs at $\mu=1.2p_{\rm T}$. The difference between $\hat{q}_0=1.2 \sim 1.4$ GeV$^2/$fm is narrow relative to the same scale case.

\begin{widetext}
\begin{center}
\begin{figure}[htbp]
\centering
\includegraphics[width=0.32\textwidth]{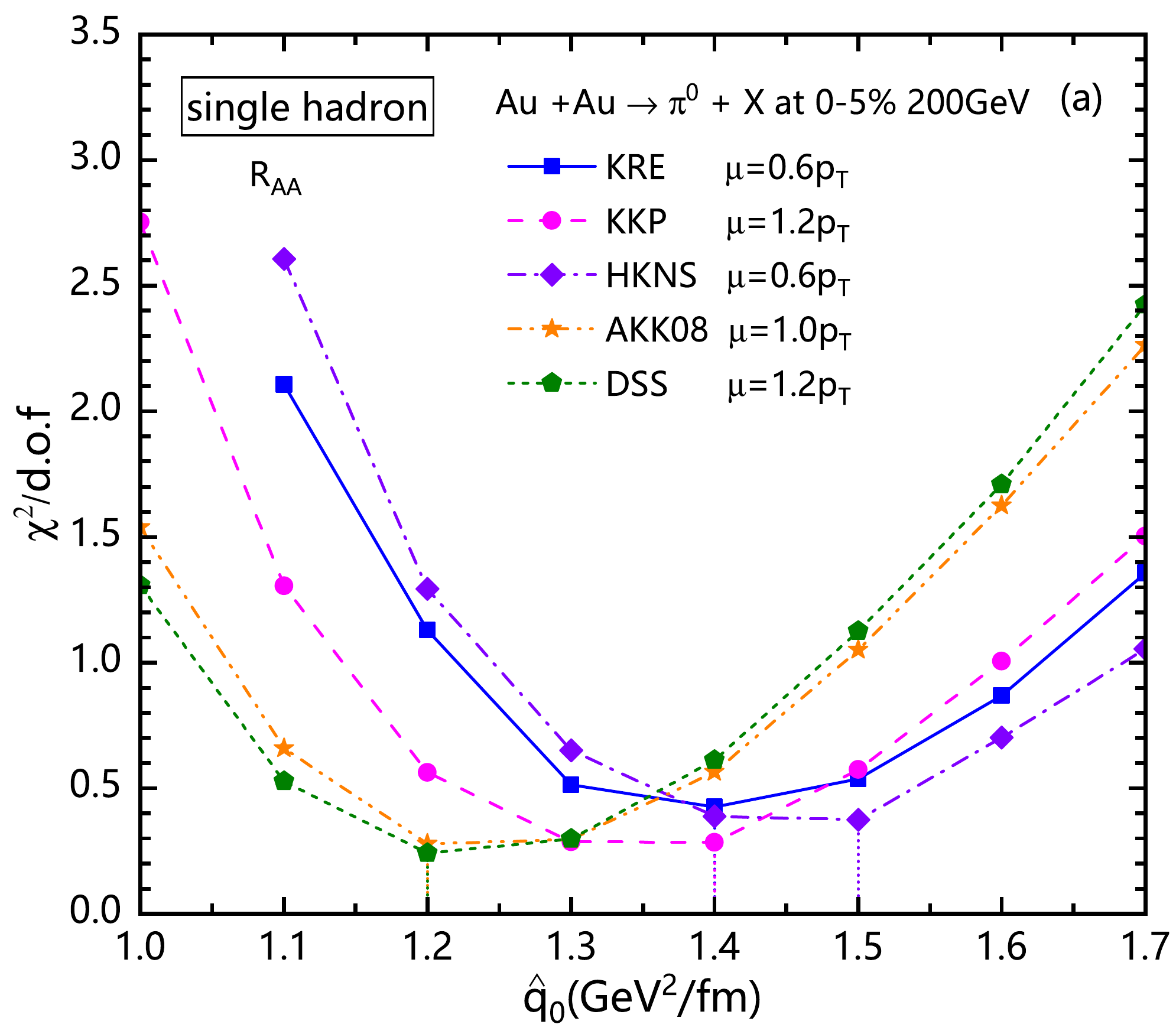}
\includegraphics[width=0.32\textwidth]{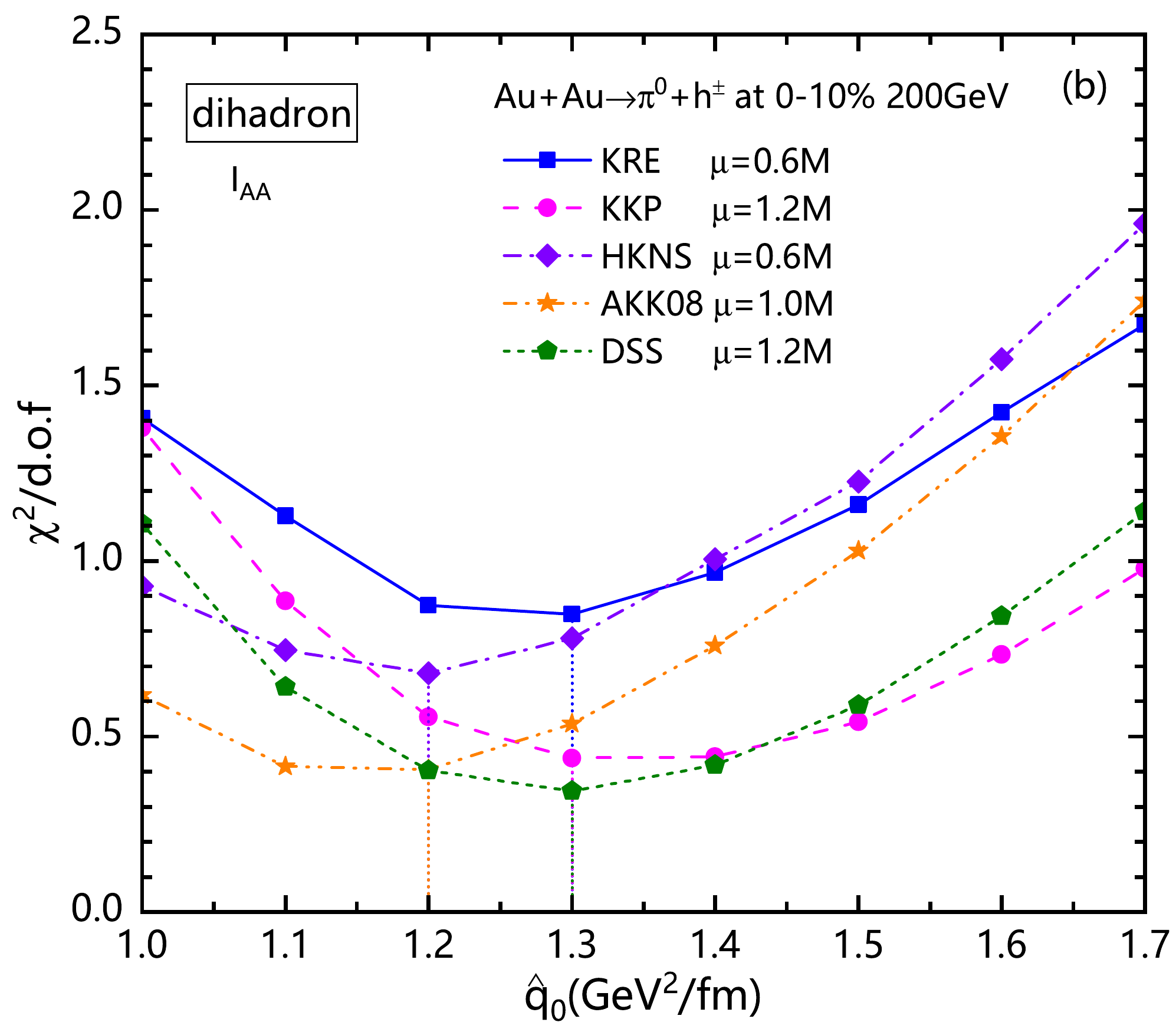}
\includegraphics[width=0.32\textwidth]{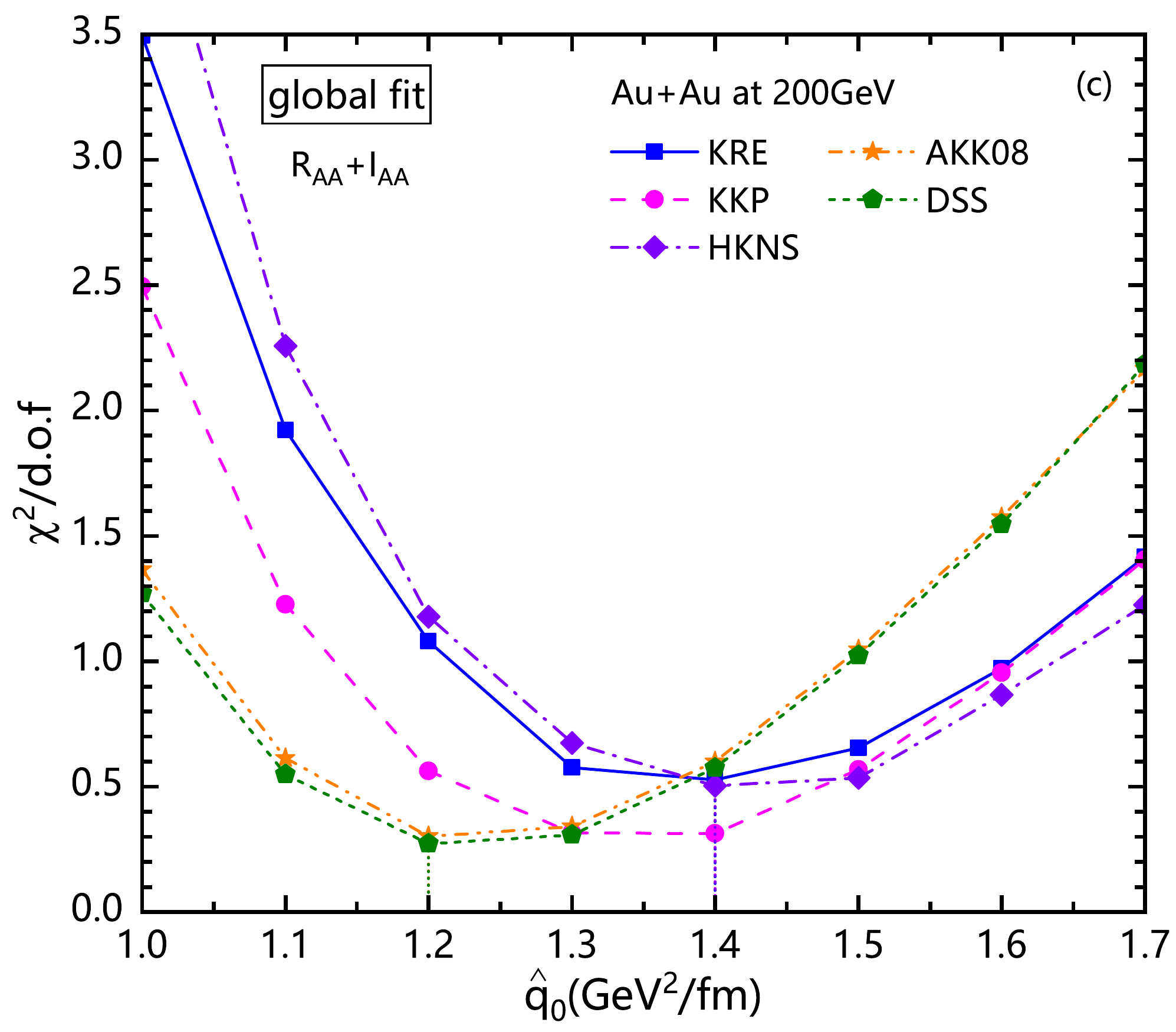}
\caption{The $\chi^2/d.o.f$ results from fitting to nuclear modification factors in central $Au+Au$ collisions at $\sqrt{s_{\rm NN}}=200$ GeV, (a) the fits to only single hadron $R_{AA}(p_{\rm T})$, (b) the fits to only dihadron $I_{AA}(z_{\rm T})$, and (c) the global fits to $R_{AA}(p_{\rm T})+I_{AA}(z_{\rm T})$. 5 sets of fragmentation function parameterizations are used in the model with different scales, respectively.}
\label{fig:kai2_rhic_d}
\end{figure}
\end{center}
\begin{center}
\begin{figure}[htbp]
\includegraphics[width=0.32\textwidth]{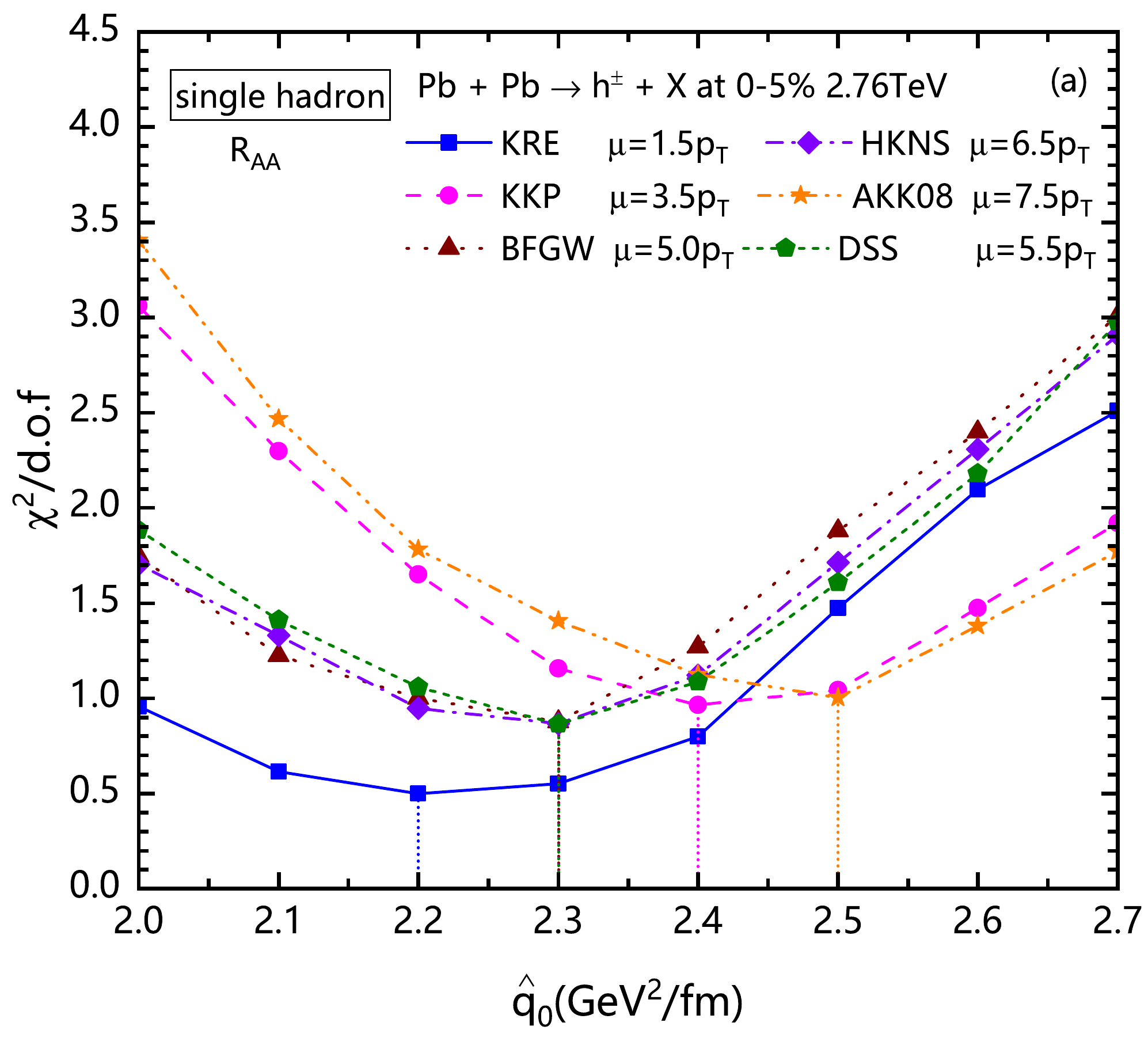}
\includegraphics[width=0.32\textwidth]{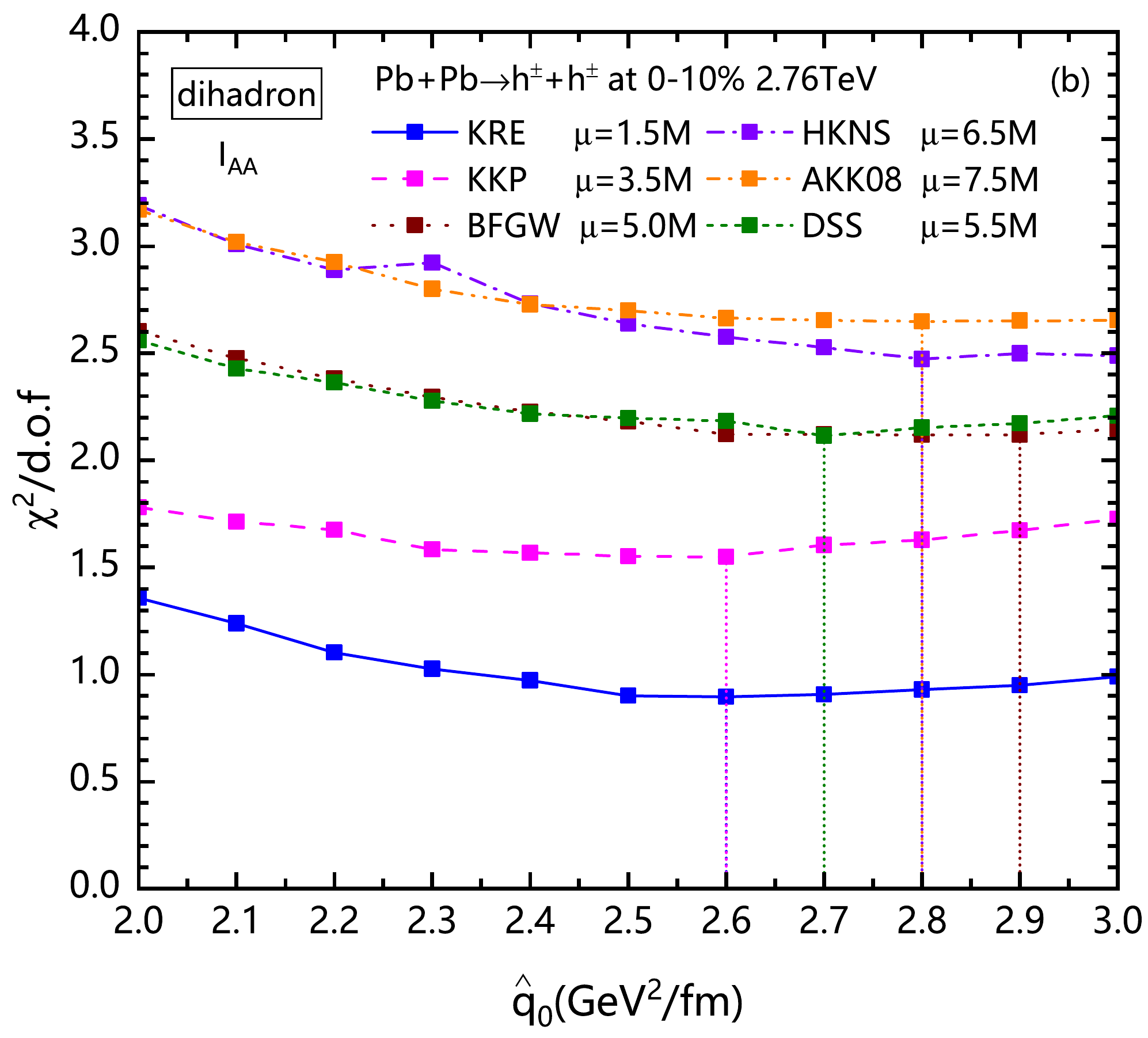}
\includegraphics[width=0.32\textwidth]{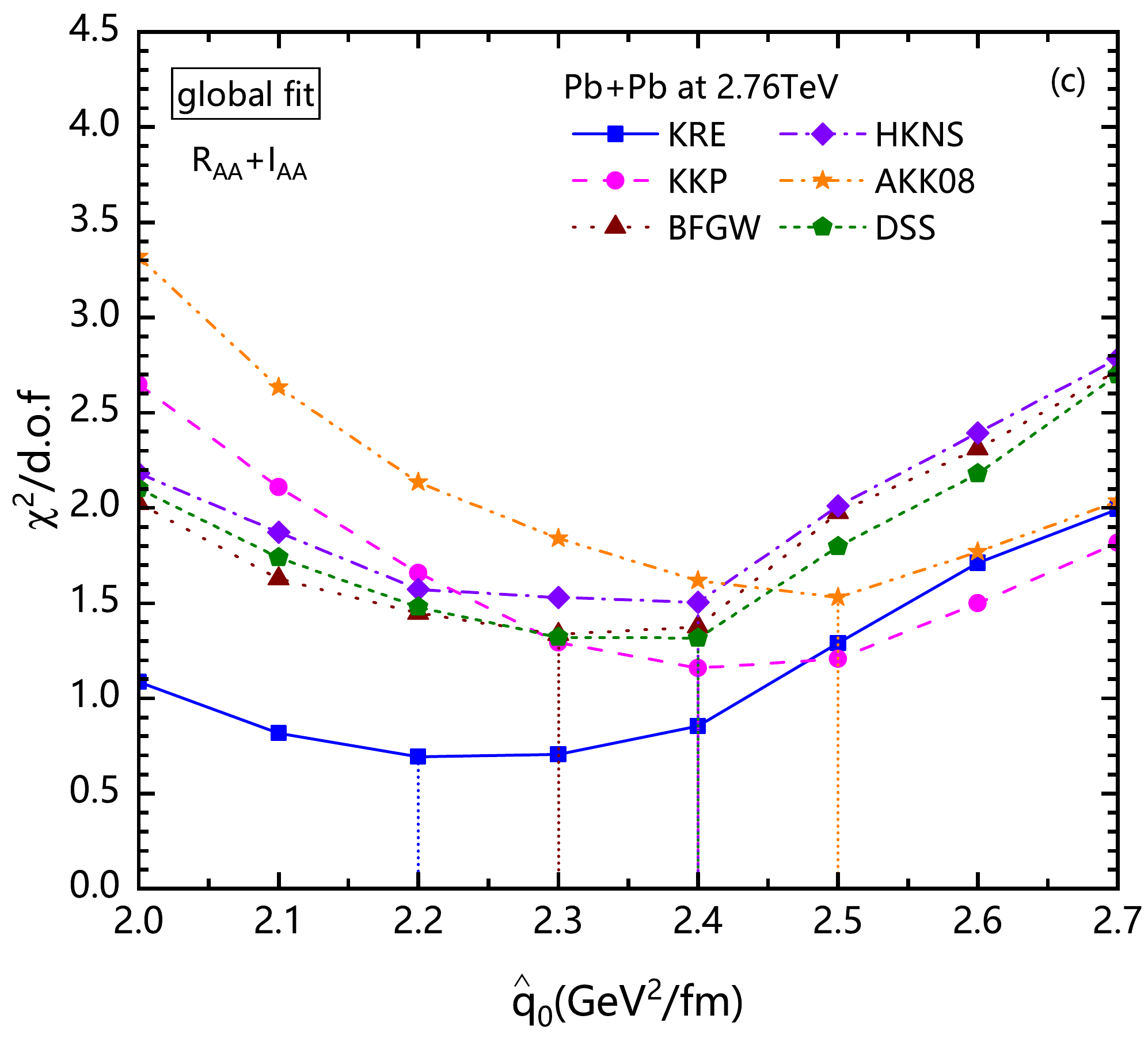}
\caption{The $\chi^2/d.o.f$ results from fitting to nuclear modification factors in central $Pb+Pb$ collisions at $\sqrt{s_{\rm NN}}=2.76$ TeV, (a) the fitts to only single hadron $R_{AA}(p_{\rm T})$  \cite{Abelev:2012hxa, CMS:2012aa}, (b) the fits to only dihadron $I_{AA}(p_{\rm T}^{\rm assoc})$ \cite{ALICE:2011gpa, Conway:2013xaa}, and (c) the global fits to $R_{AA}(p_{\rm T})+I_{AA}(p_{\rm T}^{\rm assoc})$. 6 sets of fragmentation function parameterizations are used in the model with different scales, respectively.}
\label{fig:kai2_lhc_d}
\end{figure}
\end{center}
\end{widetext}

Fig.~\ref{fig:kai2_lhc_d} shows the $\chi^2/d.o.f$ fits to nuclear modification factors in central $Pb+Pb$ collisions at $\sqrt{s_{\rm NN}}=2.76$ TeV: (a) the fits to only single hadron $R_{AA}(p_{\rm T})$, (b) the fits to only dihadron $I_{AA}(p_{\rm T}^{\rm assoc})$, and (c) the global fits to $R_{AA}(p_{\rm T})$ + $I_{AA}(p_{\rm T}^{\rm assoc})$. We use 6 sets of fragmentation function parameterizations with different scales in our calculations, respectively.
We get the best fitting values of the jet transport parameter as: $\hat{q}_0=2.2$ GeV$^2/$fm with KRE FFs at $\mu=1.5p_{\rm T}$, $\hat{q}_0=2.4$ GeV$^2/$fm with KKP FFs at $\mu=3.5p_{\rm T}$, $\hat{q}_0=2.3$ GeV$^2/$fm with BFGW FFs at $\mu=5.0p_{\rm T}$, $\hat{q}_0=2.4$ GeV$^2/$fm with HKNS FFs at $\mu=6.5p_{\rm T}$, $\hat{q}_0=2.5$ GeV$^2/$fm with AKK08 FFs at $\mu=7.5p_{\rm T}$, $\hat{q}_0=2.4$ GeV$^2/$fm with DSS FFs at $\mu=5.5p_{\rm T}$. The difference between $\hat{q}_0$ drops to $2.2\sim2.5$ GeV$^2/$fm.

The above transport coefficients $\hat{q}/T^3$ extracted in the model with different scales in different sets of FFs from hadron suppressions at RHIC and the LHC are summarized in the right panel of Fig. \ref{fig:qhat_uncertainty}.

\subsection{Characteristics of fragmentation function contributions for parton to hadron with different scales for different sets of FFs}

Shown in Fig.~\ref{fig:zD-z-pi_rhic_d} are the quark and gluon FFs of $\pi^0$ hadrons at $p_{\rm T}=10$ GeV and 50 GeV for all the available FFs with different scales, respectively. Fig.~\ref{fig:zD-z-h_lhc_d} are the similar plot for charged-hadron FFs of quark and gluon with different scales in each FFs. The difference between gluon FFs are a bit reduced with befitted scales compared with the same scale case.

\begin{center}
\begin{figure}[htbp]
\includegraphics[width=0.40\textwidth]{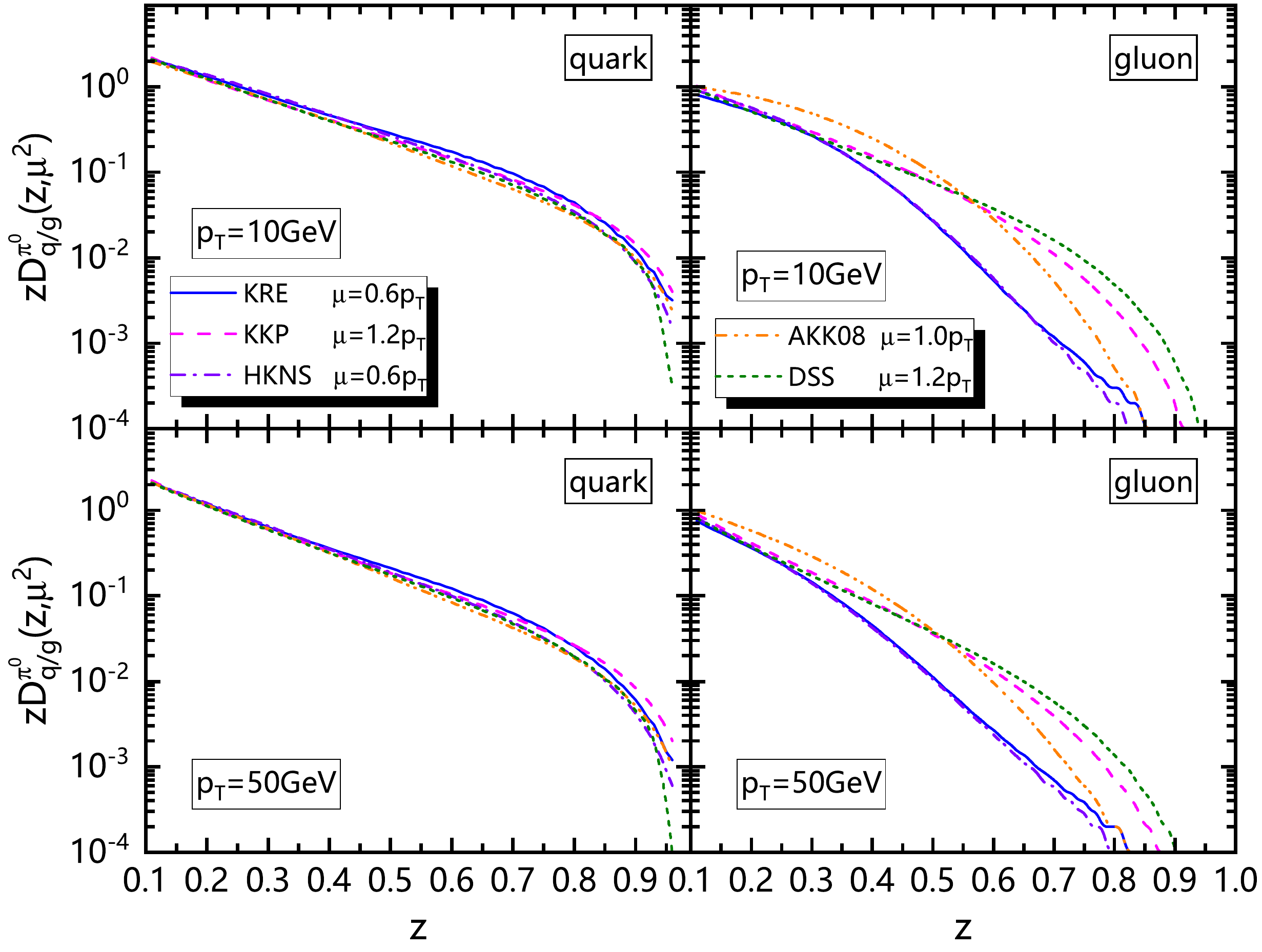}
\caption{Comparisons of parton-to-$\pi^0$ fragmentation functions between 5 sets of FFs: left panels are for quark FFs, and right panels for gluon FFs. Upper panels  are for the hadrons with $p_{\rm T}=10$ GeV, and lower panels with $p_{\rm T}=50$ GeV. The scale $\mu$ is different for each FFs.}
\label{fig:zD-z-pi_rhic_d}
\end{figure}
\begin{figure}[htbp]
\includegraphics[width=0.40\textwidth]{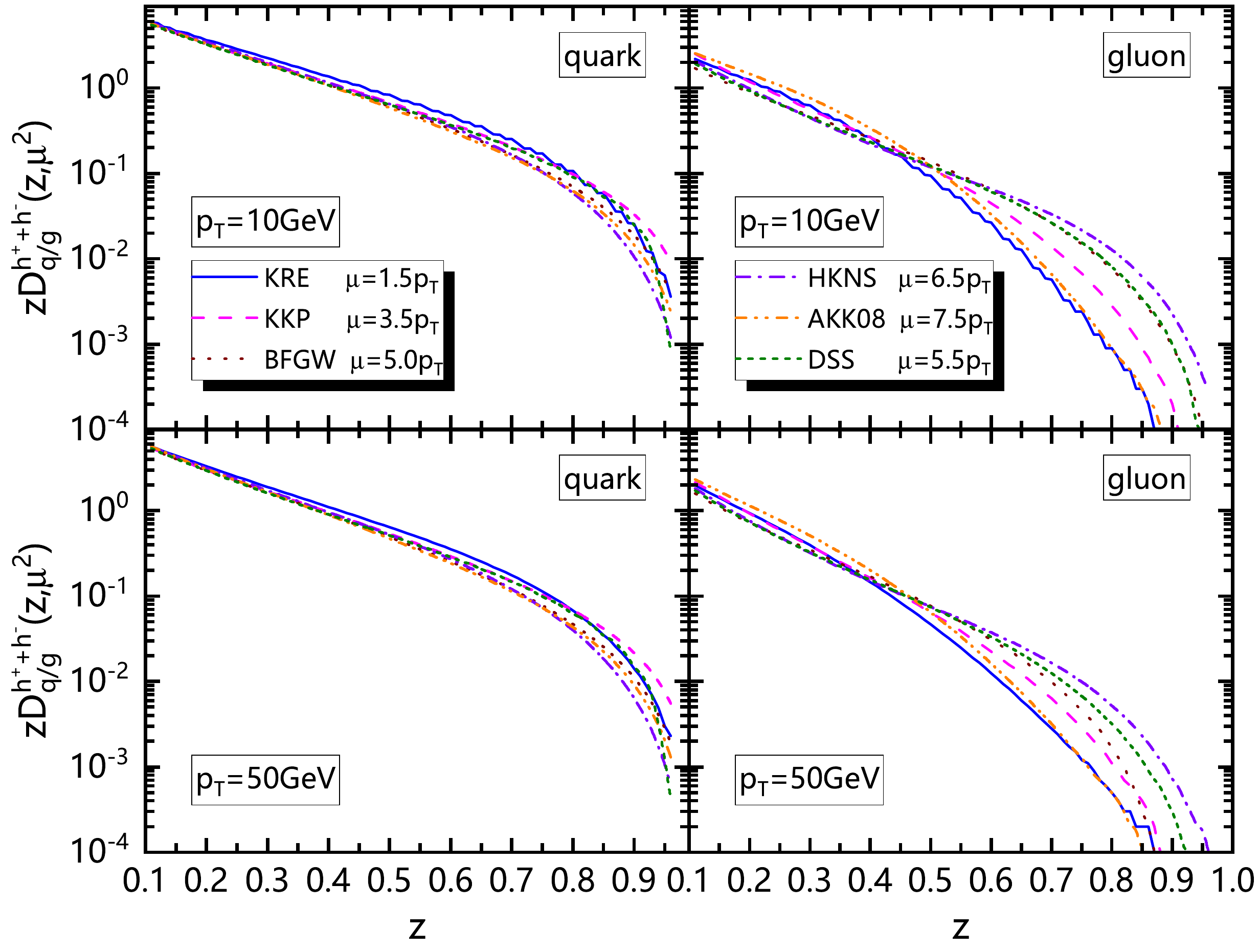}
\caption{Comparisons of parton-to-$h^{\pm}$ FFs between 6 sets of FFs: left panels are for quark FFs, and right panels for gluon FFs. Upper panels are for the hadrons with $p_{\rm T}=10$ GeV, and lower panels with $p_{\rm T}=50$ GeV. The scale $\mu$ is different for each FFs.}
\label{fig:zD-z-h_lhc_d}
\end{figure}
\end{center}

\begin{figure}[htbp]
\includegraphics[width=0.40\textwidth]{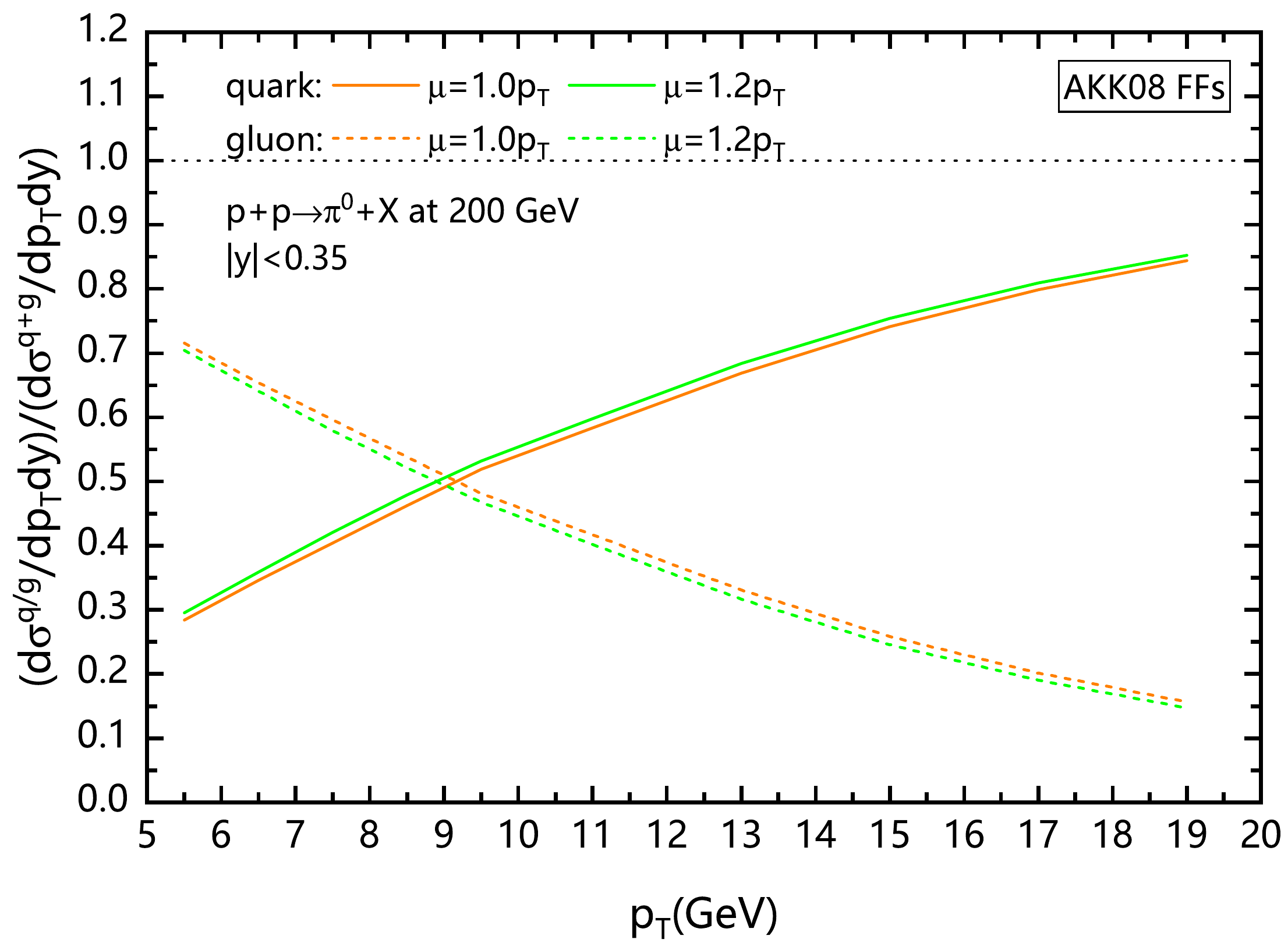}
\caption{The contribution fractions of quark (solid lines) and gluon (dashed lines) fragmentations to the inclusive $\pi^0$ cross sections given by the model with only AKK08 FFs with $\mu=1.0p_{\rm T}$ and $\mu=1.2p_{\rm T}$ at $\sqrt{s_{\rm NN}}=200$ GeV, respectively.}
\label{fig:frac_akk08_rhic}
\end{figure}

\begin{figure}[htbp]
\includegraphics[width=0.40\textwidth]{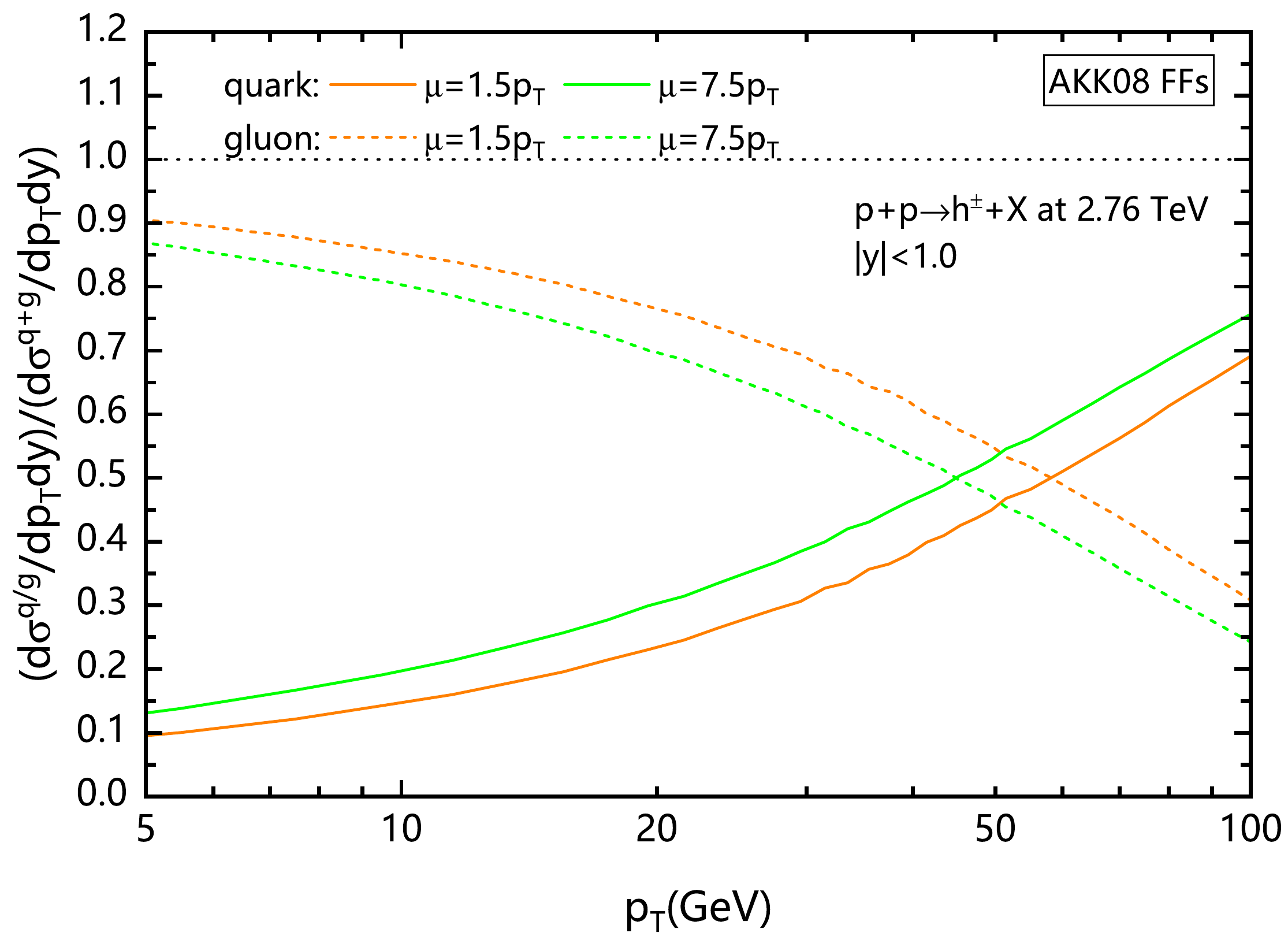}
\caption{The contribution fractions of quark (solid lines) and gluon (dashed lines) fragmentations to the inclusive charged-hadron cross sections given by the model with only AKK08 FFs with $\mu=1.5p_{\rm T}$ and $\mu=7.5p_{\rm T}$ at $\sqrt{s_{\rm NN}}=2.76$ TeV, respectively.}
\label{fig:frac_akk08_lhc}
\end{figure}

Taking the AKK08 FFs as examples, we show the relative contributions of quark and gluon to hadrons with the changed scales in $p+p$ collisions at RHIC and the LHC energies, as shown in Fig. \ref{fig:frac_akk08_rhic} and \ref{fig:frac_akk08_lhc}. One can see that both at RHIC and the LHC, with $\mu$ increasing, the gluon contribution to final state hadrons will decrease, thus we need a larger jet quenching parameter to compensate for the total jet energy loss.

For a given set of FFs in a NLO pQCD parton model, the different scales leads to the different fractions of gluon (quark) contributions to hadrons in $p+p$ collisions. Although the scale change can also affect the parton distribution functions and the hard cross sections, the fraction change is mainly contributed by parton fragmentation functions, as shown in Fig. \ref{fig:frac_akk08_rhic} and \ref{fig:frac_akk08_lhc}. In details, when the scale $\mu$ decreases from 1.2$p_{\rm T}$ to 1.0$p_{\rm T}$ at $\sqrt{s_{\rm NN}}=200$ GeV, the contribution of gluon-to-hadron becomes larger, as illustrated in Fig. \ref{fig:frac_akk08_rhic}, so a relatively smaller $\hat{q}_0$ is needed for the case of $\mu = 1.0p_{\rm T}$ shown in Fig. \ref{fig:qhat_uncertainty}. Meanwhile, with the $\mu$ increasing from 1.5$p_{\rm T}$ to 7.5$p_{\rm T}$ at $\sqrt{s_{\rm NN}}=2.76$ TeV, the contribution of gluon-to-hadron reduces, as presented in Fig. \ref{fig:frac_akk08_lhc}, thus a relatively larger $\hat{q}_0$ is needed for the case of $\mu = 7.5p_{\rm T}$ shown in Fig. \ref{fig:qhat_uncertainty}. In a word, the different fraction of gluon (quark) contribution to hadrons will give different energy loss parameters due to gluon energy loss being $9/4$ times of quark energy loss.

\bibliography{qhat-nFFs}

\end{document}